\documentclass[twocolumn]{aastex62}

\usepackage{gensymb}

\graphicspath{{./}{figures/}}

\usepackage{graphicx}

\received{\today}
\revised{...}
\accepted{...}
\submitjournal{ApJ}

\shorttitle{VERITAS VHE Discovery of 3C\,264}
\shortauthors{A.~Archer, et al.}
\newcommand{\be}{\begin{itemize}}
\newcommand{\ee}{\end{itemize}}

\newcommand{\Jybm}{\hbox{${\rm Jy \ beam}^{-1}\;$}}

\begin{document}

\title{VERITAS Discovery of VHE Emission from the Radio Galaxy 3C 264: A Multi-Wavelength Study}

\correspondingauthor{Eileen T. Meyer}
\email{meyer@umbc.edu}

\correspondingauthor{Wystan Benbow}
\email{wbenbow@cfa.harvard.edu}

\correspondingauthor{Jodi Christiansen}
\email{jlchrist@calpoly.edu}

\correspondingauthor{Lucy Fortson}
\email{lfortson@umn.edu}

\correspondingauthor{Markos Georganopoulos}
\email{georgano@umbc.edu}

% Matt says it is not important to be a corresponding author if we run into issues.
\correspondingauthor{Matt Lister}
\email{mlister@purdue.edu}

% Authors included on March 3 ~ 1 pm;  Through Archer sign-up

% Some associates need manual additions
% Lister \author{M.~L.~Lister}\affiliation{Department of Physics and Astronomy, Purdue University, 525 Northwestern Avenue, West Lafayette, IN 47907, USA}  -- Watch out for other Purdue affiliations
% Georganopoulos \author{M.~Georganopoulos}\affiliation{Department of Physics, University of Maryland, Baltimore County, Baltimore MD 21250, USA}  -- matches Eileen
% Alberto  \author{A.~Sadun}\affiliation{Department of Physics $|$ University of Colorado Denver, Denver, CO 80217, USA}
%  Teddy Cheung & Tyrel Johnson as they declined authorship (complexity of CAT-2 review issues).
% M. Carini \author{M.T. Carini}\affiliation{Department of Physics and Astronomy $|$ Western Kentucky University, Bowling Green, KY 42103, USA}

\author{A.~Archer}\affiliation{Department of Physics and Astronomy, DePauw University, Greencastle, IN 46135-0037, USA}
\author{W.~Benbow}\affiliation{Center for Astrophysics $|$ Harvard \& Smithsonian, Cambridge, MA 02138, USA}
\author{R.~Bird}\affiliation{Department of Physics and Astronomy, University of California, Los Angeles, CA 90095, USA}
\author{A.~Brill}\affiliation{Physics Department, Columbia University, New York, NY 10027, USA}
\author{M.~Buchovecky}\affiliation{Department of Physics and Astronomy, University of California, Los Angeles, CA 90095, USA}
\author{J.~H.~Buckley}\affiliation{Department of Physics, Washington University, St. Louis, MO 63130, USA}
\author{M.T. Carini}\affiliation{Department of Physics and Astronomy $|$ Western Kentucky University, Bowling Green, KY 
42103, USA}
\author{J.~L.~Christiansen}\affiliation{Physics Department, California Polytechnic State University, San Luis Obispo, CA 94307, USA}
\author{A.~J.~Chromey}\affiliation{Department of Physics and Astronomy, Iowa State University, Ames, IA 50011, USA}
\author{M.~K.~Daniel}\affiliation{Center for Astrophysics $|$ Harvard \& Smithsonian, Cambridge, MA 02138, USA}
\author{M.~Errando}\affiliation{Department of Physics, Washington University, St. Louis, MO 63130, USA}
\author{A.~Falcone}\affiliation{Department of Astronomy and Astrophysics, 525 Davey Lab, Pennsylvania State University, University Park, PA 16802, USA}
\author{Q.~Feng}\affiliation{Physics Department, Columbia University, New York, NY 10027, USA}
\author{P.~Fortin}\affiliation{Center for Astrophysics $|$ Harvard \& Smithsonian, Cambridge, MA 02138, USA}
\author{L.~Fortson}\affiliation{School of Physics and Astronomy, University of Minnesota, Minneapolis, MN 55455, USA}
\author{A.~Furniss}\affiliation{Department of Physics, California State University - East Bay, Hayward, CA 94542, USA}
\author{A.~Gent}\affiliation{School of Physics and Center for Relativistic Astrophysics, Georgia Institute of Technology, 837 State Street NW, Atlanta, GA 30332-0430}
\author{M.~Georganopoulos}\affiliation{Department of Physics, University of Maryland, Baltimore County, Baltimore MD 21250, USA}
\author{G.~H.~Gillanders}\affiliation{School of Physics, National University of Ireland Galway, University Road, Galway, Ireland}
\author{C.~Giuri}\affiliation{DESY, Platanenallee 6, 15738 Zeuthen, Germany}
\author{O.~Gueta}\affiliation{DESY, Platanenallee 6, 15738 Zeuthen, Germany}
\author{D.~Hanna}\affiliation{Physics Department, McGill University, Montreal, QC H3A 2T8, Canada}
\author{T.~Hassan}\affiliation{DESY, Platanenallee 6, 15738 Zeuthen, Germany}
\author{O.~Hervet}\affiliation{Santa Cruz Institute for Particle Physics and Department of Physics, University of California, Santa Cruz, CA 95064, USA}
\author{J.~Holder}\affiliation{Department of Physics and Astronomy and the Bartol Research Institute, University of Delaware, Newark, DE 19716, USA}
\author{G.~Hughes}\affiliation{Center for Astrophysics $|$ Harvard \& Smithsonian, Cambridge, MA 02138, USA}
\author{T.~B.~Humensky}\affiliation{Physics Department, Columbia University, New York, NY 10027, USA}
\author{P.~Kaaret}\affiliation{Department of Physics and Astronomy, University of Iowa, Van Allen Hall, Iowa City, IA 52242, USA}
\author{M.~Kertzman}\affiliation{Department of Physics and Astronomy, DePauw University, Greencastle, IN 46135-0037, USA}
\author{D.~Kieda}\affiliation{Department of Physics and Astronomy, University of Utah, Salt Lake City, UT 84112, USA}
\author{F.~Krennrich}\affiliation{Department of Physics and Astronomy, Iowa State University, Ames, IA 50011, USA}
\author{M.~J.~Lang}\affiliation{School of Physics, National University of Ireland Galway, University Road, Galway, Ireland}
\author{T.~T.Y.~Lin}\affiliation{Physics Department, McGill University, Montreal, QC H3A 2T8, Canada}
\author{M.~L.~Lister}\affiliation{Department of Physics and Astronomy, Purdue University, 525 Northwestern Avenue,
West Lafayette, IN 47907, USA}
\author{M.~Lundy}\affiliation{Physics Department, McGill University, Montreal, QC H3A 2T8, Canada}
\author{G.~Maier}\affiliation{DESY, Platanenallee 6, 15738 Zeuthen, Germany}
\author{E.~T.~Meyer}\affiliation{Department of Physics, University of Maryland, Baltimore County, Baltimore MD 21250, USA}
\author{P.~Moriarty}\affiliation{School of Physics, National University of Ireland Galway, University Road, Galway, Ireland}
\author{R.~Mukherjee}\affiliation{Department of Physics and Astronomy, Barnard College, Columbia University, NY 10027, USA}
\author{D.~Nieto}\affiliation{Institute of Particle and Cosmos Physics, Universidad Complutense de Madrid, 28040 Madrid, Spain}
\author{M.~Nievas-Rosillo}\affiliation{DESY, Platanenallee 6, 15738 Zeuthen, Germany}
\author{S.~O'Brien}\affiliation{Physics Department, McGill University, Montreal, QC H3A 2T8, Canada}
\author{R.~A.~Ong}\affiliation{Department of Physics and Astronomy, University of California, Los Angeles, CA 90095, USA}
\author{K.~Pfrang}\affiliation{DESY, Platanenallee 6, 15738 Zeuthen, Germany}
\author{M.~Pohl}\affiliation{Institute of Physics and Astronomy, University of Potsdam, 14476 Potsdam-Golm, Germany and DESY, Platanenallee 6, 15738 Zeuthen, Germany}
\author{R.~R.~Prado}\affiliation{DESY, Platanenallee 6, 15738 Zeuthen, Germany}
\author{E.~Pueschel}\affiliation{DESY, Platanenallee 6, 15738 Zeuthen, Germany}
\author{J.~Quinn}\affiliation{School of Physics, University College Dublin, Belfield, Dublin 4, Ireland}
\author{K.~Ragan}\affiliation{Physics Department, McGill University, Montreal, QC H3A 2T8, Canada}
\author{K.~Ramirez}\affiliation{Physics Department, California Polytechnic State University, San Luis Obispo, CA 94307, USA}
\author{P.~T.~Reynolds}\affiliation{Department of Physical Sciences, Cork Institute of Technology, Bishopstown, Cork, Ireland}
\author{D.~Ribeiro}\affiliation{Physics Department, Columbia University, New York, NY 10027, USA}
\author{G.~T.~Richards}\affiliation{Department of Physics and Astronomy and the Bartol Research Institute, University of Delaware, Newark, DE 19716, USA}
\author{E.~Roache}\affiliation{Center for Astrophysics $|$ Harvard \& Smithsonian, Cambridge, MA 02138, USA}
\author{C.~Rulten}\affiliation{School of Physics and Astronomy, University of Minnesota, Minneapolis, MN 55455, USA}
\author{J.~L.~Ryan}\affiliation{Department of Physics and Astronomy, University of California, Los Angeles, CA 90095, USA}
\author{A.~Sadun}\affiliation{Department of Physics $|$ University of Colorado Denver, Denver, CO 80217, USA}
\author{M.~Santander}\affiliation{Department of Physics and Astronomy, University of Alabama, Tuscaloosa, AL 35487, USA}
\author{S.~S.~Scott}\affiliation{Santa Cruz Institute for Particle Physics and Department of Physics, University of California, Santa Cruz, CA 95064, USA}
\author{G.~H.~Sembroski}\affiliation{Department of Physics and Astronomy, Purdue University, West Lafayette, IN 47907, USA}
\author{K.~Shahinyan}\affiliation{School of Physics and Astronomy, University of Minnesota, Minneapolis, MN 55455, USA}
\author{R.~Shang}\affiliation{Department of Physics and Astronomy, University of California, Los Angeles, CA 90095, USA}
\author{B.~Stevenson}\affiliation{Department of Physics and Astronomy, University of California, Los Angeles, CA 90095, USA}
\author{V.~V.~Vassiliev}\affiliation{Department of Physics and Astronomy, University of California, Los Angeles, CA 90095, USA}
\author{S.~P.~Wakely}\affiliation{Enrico Fermi Institute, University of Chicago, Chicago, IL 60637, USA}
\author{A.~Weinstein}\affiliation{Department of Physics and Astronomy, Iowa State University, Ames, IA 50011, USA}
\author{P.~Wilcox}\affiliation{School of Physics and Astronomy, University of Minnesota, Minneapolis, MN 55455, USA}
\author{A.~Wilhelm}\affiliation{Institute of Physics and Astronomy, University of Potsdam, 14476 Potsdam-Golm, Germany and DESY, Platanenallee 6, 15738 Zeuthen, Germany}
\author{D.~A.~Williams}\affiliation{Santa Cruz Institute for Particle Physics and Department of Physics, University of California, Santa Cruz, CA 95064, USA}
\author{T.~J~Williamson}\affiliation{Department of Physics and Astronomy and the Bartol Research Institute, University of Delaware, Newark, DE 19716, USA}



\begin{abstract}
%The FR-I radio galaxy 3C\,264, also known as NGC\,3862, 
The radio source 3C\,264, hosted by the giant elliptical galaxy NGC\,3862, was observed with VERITAS between February 2017 and May 2019.  These deep observations resulted in the discovery of very-high-energy (VHE; E $>100$ GeV) $\gamma$-ray emission from this active galaxy. An analysis of $\sim$57 hours of quality-selected live time yields
a detection at the position of the source, 
corresponding to a statistical significance of 7.8 standard deviations above background. The observed VHE flux is variable on monthly time scales, with an elevated flux 
seen in 2018 observations. The VHE emission during this elevated state is well-characterized by a power-law spectrum with a photon index $\Gamma = 2.20 \pm 0.27$ and flux F($>315$ GeV) = ($7.6\pm 1.2_{\mathrm stat} \pm 2.3_{\mathrm syst})\times 10^{-13}$ cm$^{-2}$ s$^{-1}$, or approximately 0.7\% of the Crab Nebula flux above the same threshold. 3C\,264 ($z = 0.0217$) is the most distant radio galaxy detected at VHE, and the elevated state 
is thought to be similar to that of the famously outbursting jet in M\,87.
Consequently, extensive contemporaneous multi-wavelength data were acquired in 2018 at the time of the VHE high state.
An analysis of these data, including VLBA, VLA, HST, \emph{Chandra} and \emph{Swift} observations in addition to the VERITAS data, is presented, along with a discussion of the resulting spectral energy distribution. 

\end{abstract}



\section{Introduction} \label{sec:intro}

Active galactic nuclei (AGN) constitute a small fraction of super-massive black holes at the centers of galaxies. These objects are powered by accretion and a further fraction ($\sim$10\%) of AGN have highly collimated jets of fully ionized plasma that can reach scales of several Mpc. Numerous AGN are known to emit high-energy (HE; MeV$-$GeV) 
and very-high-energy (VHE; E $>100$ GeV) $\gamma$-rays, presumably via inverse-Compton (IC) emission of leptonic particles within the jet. All but four of the 78 (jetted) AGN currently detected at VHE are blazars, where the jet is viewed nearly along its axis; the other four are radio galaxies where the jet associated with the AGN is viewed at somewhat larger angles.\footnote{TeVCat online source catalog: \citet{wakely2008}} 
It has been suggested that radio galaxies form the parent population of blazars with core dominated objects (FR-Is, after the \cite{Fanaroff1974} classification) corresponding to BL-Lac objects observed at larger jet viewing angles, while lobe-dominated FR-II radio galaxies are, instead, associated with flat-spectrum radio quasars (FSRQ) \citep{Urry1995}.

The jet emission in blazars and radio galaxies is characterized by a double-peaked, non-thermal spectral energy distribution (SED). The lower frequency peak, which in blazars has a peak frequency ($\nu_{\mathrm peak}$) ranging from $10^{13}$ to $10^{18}$~Hz, is well-described as synchrotron emission from relativistic electrons spiraling in the magnetic field of the jet. The higher frequency peak, located in the $\gamma$-ray band,  is generally attributed to inverse-Compton (IC) emission.
Typically, the sources with higher-frequency synchrotron peaks have higher-frequency IC peaks (i.e.\ beyond the 10$-$100 GeV range).  Correspondingly, high-synchrotron-peaked (HSP) blazars ($\nu_{\mathrm peak}$ $>10^{15}$~Hz) are the brightest and best-studied AGN at VHE (51 of the current VHE AGN), even though they are the least luminous / powerful. In contrast, only 9 blazars (BL\,Lac objects and quasars) in the current TeV catalog are low-synchrotron-peaked (LSP; $\nu_{\mathrm peak}$ $\lesssim 10^{14}$~Hz) objects. Although there is no  strict division between the classes, radio galaxies are believed to have their jets oriented at larger angles to the line-of-sight ($\gtrsim$10$^\circ$) than blazars. This larger misalignment means that radio galaxies are much less Doppler boosted than their blazar counterparts, and that they tend to have lower-frequency synchrotron peaks, similar to LSP blazars \citep{meyer2011}. 

The IC emission in blazars and radio galaxies can arise from synchrotron self-Compton (SSC) or external Compton processes, or a combination of the two. It is generally thought that most low-power blazars (i.e.\ HSPs) have IC peaks dominated by SSC emission \cite[e.g.][]{boettcher2007,paggi2009} while the more powerful blazars (i.e.\ LSPs) are likely to require external-Compton processes \citep{sikora2009,Meyer2012-EC}. In the latter case, it is unclear which external photon field provides the seed photons for scattering, due to the uncertainty in the actual location of the high-energy emitting region in the jet \citep[e.g.][]{Arsioli2018}.  The possibilities for the dominant seed-photon source include the molecular torus region, the much smaller broad-line emitting region, and even the accretion disk \citep{dermer1992,sikora1994,Blazejowski2000,sikora2009}. In addition to these purely leptonic scenarios, there are also models for jet emission which include a significant population of relativistic protons (i.e.\ hadronic models) that produce 
HE and VHE $\gamma$-ray emission via several different processes \citep[e.g.][]{aharonian2000}. In particular, cloud-jet interaction models could explain the observed TeV flaring emission in sources like M87 \citep[e.g.][]{barkov2012}. 

This paper describes the discovery by VERITAS in VHE $\gamma$-rays of the FR-I radio galaxy 3C\,264. It is the fourth radio galaxy detected at VHE, and the most distant, at a comoving distance of 93 Mpc. All four VHE radio galaxies are low-power, with FR-I type jets.  The other VHE detections are Centaurus\,A, M\,87, and NGC\,1275 \citep[at a distance of 3.8, 16.7, and 62.5 Mpc, respectively;][]{harris2010_cena,blakeslee2009,ngcdist}.  Two of the four VHE radio galaxies show superluminal motions on kpc scales (3C\,264 and M\,87). It is plausible that some very nearby radio galaxies are designated such because their
proximity makes the identification of their host galaxy easier. At much larger distances, the same objects would likely be classified as (slightly misaligned) blazars. Indeed, the VHE source IC\,310, detected by 
both VERITAS and MAGIC \citep{Aleksic2014a}, is considered by some to be a fifth VHE radio galaxy \citep{IC310_RG1,IC310_RG2}. However, 
there are convincing arguments  that it is a borderline BL\,Lac object \citep{kadler2012}. A similar case is the VHE source PKS\,0625$-$354, for which there is also some ambiguity, although the balance of evidence is in favor of a BL Lac classification \citep{hess2018}.

Previous VHE detections of radio galaxies reveal high-energy Compton components similar to blazars in terms of spectral shape and origin, though at lower luminosity due to the decreased Doppler boosting. As is sometimes the case for blazars, single-zone SSC models are usually inadequate to explain the observed emission. In the blazar/radio galaxy IC\,310, a rising TeV component led to suggestions for a hadronic origin, or a leptonic origin with multiple electron distributions \citep{fraija2017}. A similar spectral hardening at VHE is seen in Cen\,A \citep{Aharonian2009_cenA} and possibly M\,87 \citep{rieger2018_review}. In contrast, a single-zone SSC model is compatible with the high-energy emission and variability of NGC\,1275 \citep{Aleksic2014a}. 

Comparing jet structure within VHE radio galaxies, 3C\,264 closely resembles M\,87. Both have one-sided FR-I type jets with the same kinetic luminosity ($10^{43.8}$ erg s$^{-1}$; \citealp{meyer2011}).  Their
jets also have similar morphological traits (i.e.\ multiple knots) and they share 
similar qualitative kinematic characteristics within the jet substructure \citep{meyer2015_nature}. 
In contrast, NGC\,1275 and Cen\,A both have misaligned two-sided radio jets. 
M\,87 has famously shown an outburst in the optical and X-ray bands from HST-1, a knot $\sim$100 pc downstream of the core (sky-projected) \citep{harris2006_m87}. It has also exhibited extreme (day-scale) VHE variability on multiple occasions \citep{aharonian2006,harris2009_m87, aliu2012}, which some attribute to HST-1, rather than the core. In light of the similarities between M\,87 and 3C\,264, and due to the ongoing collision between two knots in the 3C\,264 jet \citep{meyer2015_nature},
a suite of contemporaneous multi-wavelength observations was assembled to complement the detection of an increased VHE flux from 3C\,264 in early 2018.
The goal was to observe a change in brightness or structure within the jet or core during the same period. Therefore, in addition to the VERITAS VHE discovery of 3C 264, this paper describes the results from this multi-wavelength observation campaign, 
as well as the similarities and differences between 
3C\,264 and M\,87, particularly in light of their variability on 100$-$1000 parsec scales and at VHE. 

In this paper a standard $\Lambda$CDM cosmology is assumed with $H_0$ = 67.8 km~s$^{-1}$~Mpc$^{-1}$, $\Omega_M$ = 0.308, and $\Omega_\Lambda$ = 0.692. The luminosity distances to 3C\,264 and M\,87 are 95.4 and 22.2 Mpc, respectively.

\section{VERITAS Data \& Results } \label{sec:data}

%%%%%%%%%%%%%%%%%%%%%%%%%%%%%%%%%%%%%%%%%%%%%%%%%%%%%%%
% VERITAS DATA
%%%%%%%%%%%%%%%%%%%%%%%%%%%%%%%%%%%%%%%%%%%%%%%%%%%%%%%
\label{DataVHE}

The Very Energetic Radiation Imaging Telescope Array System (VERITAS) is an array of four imaging atmospheric Cherenkov telescopes located at Fred Lawrence Whipple Observatory near Amado, Arizona (31$^\circ$ 40'N, 110$^\circ$ 57'W). The 12-m diameter telescopes are of the Davies-Cotton design, and each is instrumented with a 499 photomultiplier tube (PMT) camera providing a field-of-view of 3.5$^\circ$. The observatory is sensitive to $\gamma$-rays between $\sim$85 GeV and $\sim$30 TeV. The angular resolution of
the facility is $\sim$0.08$^\circ$ at 1 TeV, and its energy resolution is approximately 15\% \citep{Holder2006, Christiansen2017}.  

\begin{figure}[t]
  \centering
  \includegraphics[width=3.5in]{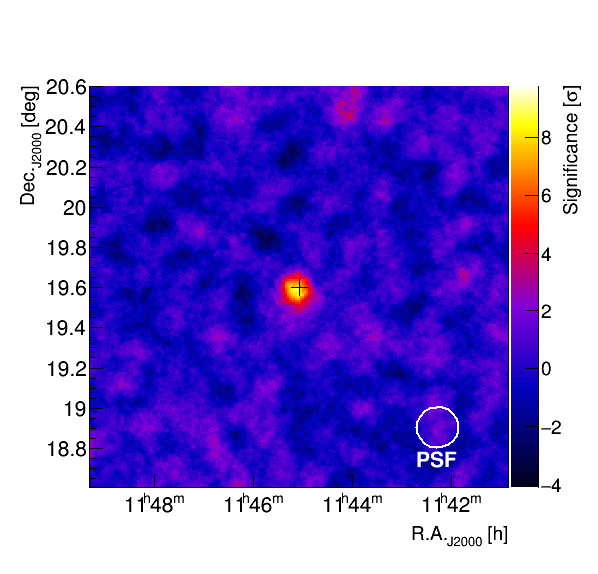}

\caption{\label{fig:VERITAS_skymap} VERITAS sky map of the significance observed from the direction of 3C\,264 during 2018. The centroid of the excess observed by VERITAS is within $2\sigma$ of the SIMBAD position of 3C\,264 (black cross). The extent of the VHE source is consistent with the VERITAS point-spread function (PSF). The PSF in this analysis is reduced from prior publications due to the use of the ITM $\gamma$-ray reconstruction \citep{Christiansen2017}.}

\end{figure}

\begin{table*}[t]
\caption{Results from VERITAS observations of 3C\,264 in 2017 $-$ 2019. The quality-selected live time, number of $\gamma$-ray-like events in the on- and off-source regions, the normalization for the larger off-source region, the observed excess of $\gamma$-rays and the corresponding statistical significance are shown.  For each observation epoch, the integral flux corresponding to the observed excess is given.  For the 2017 and 2019 observations, an upper limit may be more appropriate and this information is given in the text. The flux is reported above the observation threshold of 315 GeV, and is also given in percentage of Crab Nebula flux above the same threshold. Some quantities may not appear to sum precisely due to rounding.}
\centering
\begin{tabular}{lccccccccc}
\hline \hline
Epoch &  MJD & T      & On & Off & Norm. &  Excess & Significance & Flux ($>$315 GeV) & Crab \\
      &  & [ hr ] &    &     &       &         & [ $\sigma$ ] & [ $10^{-13}$ cm$^{-2}$ s$^{-1}$ ] & [ \% ] \\
\hline
Total (2017 $-$ 19)  & 57811$-$58633 & 57.0 & 225 & 1856  & 0.0666 & 101.4 & 7.8 & $ 5.8 \pm 0.9 $ & $0.54 \pm 0.08$ \\
\hline
Feb. $-$ May 2017   & 57811$-$57893 & 9.2  & 26  & 306  & 0.0663 & 5.7  & 1.2 & $1.9 \pm 1.7 $ & $0.18 \pm 0.16$ \\
Feb. $-$ April 2018 & 58158$-$58229 & 37.9 & 172 & 1279 & 0.0665 & 87.0 & 7.9 &  $7.6 \pm 1.2 $ & $0.71 \pm 0.11$ \\
Jan. $-$ May 2019   & 58487$-$58633 & 10.0 & 27  & 271  & 0.0674 & 8.8  & 1.8 & $2.9 \pm 1.8 $ & $0.27 \pm 0.17$\\
\hline
February 2018   & 58158$-$58170 & 3.0 & 20 & 102  & 0.0667 & 13.2 & 3.9 & $13.1 \pm 4.5$ & $1.20  \pm 0.41$ \\
March 2018      & 58186$-$58198 & 17.7 & 93 & 599 & 0.0667 & 53.0 & 6.8 & $10.2 \pm 1.9$ & $0.95 \pm 0.18$ \\
April 2018      & 58212$-$58229 & 17.2 & 59 & 578 & 0.0662 & 20.8 & 3.0 & $4.0  \pm 1.5$ & $0.37 \pm 0.14$ \\
\hline
\end{tabular}
\label{table:VERITAS_exposure}
\end{table*}

The VERITAS observations of 3C\,264 were taken from February through May 2017, from February through April 2018, and from January through May 2019.  The AGN was observed for 30 minute runs in `wobble mode' where the source position was offset from the center of the camera field of view by $0.5^{\circ}$ in each of the cardinal directions in successive runs \citep{Fomin1994}. Generally, several runs were taken on each of the individual nights during the approximately monthly `dark periods' in the three seasons of data taking.  However, the observed signal from 3C\,264 is relatively weak and therefore results are only reported for coarse temporal bins. A total of 11.0, 47.7 and 12.8 hours of data were taken in weather conditions classified as good quality by VERITAS observers in 2017, 2018 and 2019, respectively. These data are further quality selected based on information from atmospheric-monitoring instruments and the functionality of various subsystems. 

The data are reduced using the Image-Template Method (ITM) \citep{Christiansen2017}. The PSF in this analysis is reduced from prior publications due to the improved angular resolution of the ITM $\gamma$-ray reconstruction. The event-selection criteria for identifying $\gamma$-ray images and removing background cosmic-ray images is optimized for hard-spectrum sources using Crab Nebula data scaled to 1\% of its nominal strength.
The signal is extracted from a circular region of $0.0707^{\circ}$ radius centered on the International Celestial Reference Frame (ICRF) radio position of 3C~264, and the background is typically determined from 15 off-source regions with the same offset from the center of the VERITAS camera (Reflected Region Method; \cite{Berge}). The significance of any excess is calculated following Equation 17 of \cite{LiMa}.  The $\gamma$-ray selection requirements result in an average energy threshold of about 315 GeV for the conditions under which 3C\,264 was observed.

 Table~\ref{table:VERITAS_exposure} shows the results from the VERITAS observations.  Overall, an excess of 101 $\gamma$-ray-like events is observed from the direction of 3C\,264, corresponding to a statistical significance of 7.8 standard deviations ($\sigma$) above background.  While some excess is observed in both the 2017 and 2019 data sets, it is clear that a majority of the signal comes from the 2018 observations, and from February to March 2018 in particular.  
 The 2018 observations yield an excess of 87 events ($7.9 \sigma$) in 37.9 hours of live time, and the VERITAS results from these data are emphasized in this paper.  Figure~\ref{fig:VERITAS_skymap} shows the significance map for the 2018 data.  A clear point-source is seen at the position of 3C\,264. 
 
\begin{figure}[bth]
\centering
 \includegraphics[width=3.5in]{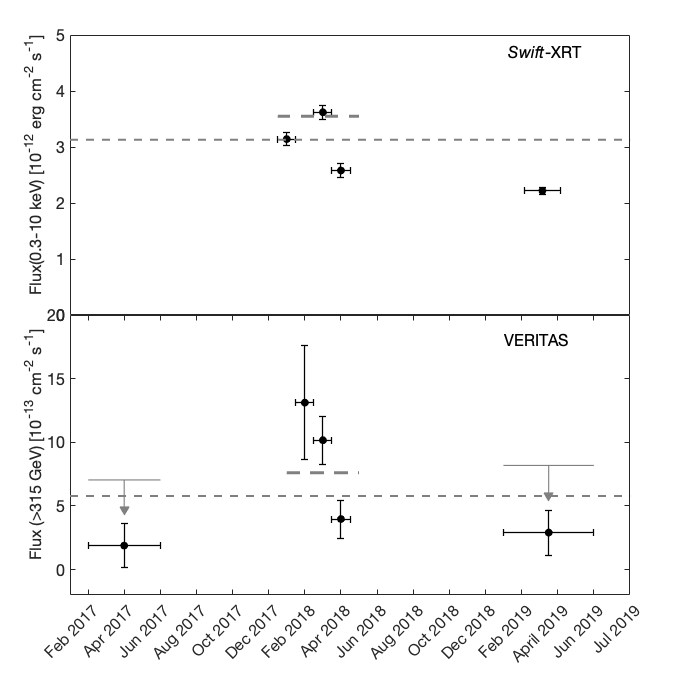}

\caption{\label{fig:VERITAS_lightcurve} \textit{Bottom: }VHE light curve measured by VERITAS.  The flux was elevated in February and March of 2018. Upper limits at the 99\% CL are also shown for the 2017 and 2019 observations due to the low significance of the observed excesses. The flux observed in 2017$-$19 is indicated by the line (short-dashed). The 2018 flux is indicated by the dashed line segment. \textit{Top:} X-ray flux light curve measured by \textit{Swift}-XRT. The average flux from 2018-2019 is indicated by a dashed line. }
\end{figure}

 The VHE light curve from 3C\,264 is shown in Figure~\ref{fig:VERITAS_lightcurve}, and all the plotted integral flux values above the observation threshold of 315 GeV are given in Table~\ref{table:VERITAS_exposure}. The systematic error on the flux measured by VERITAS is 30\%. The flux for the total 2017$-$19 measurement is shown as a line (short-dashed) in Figure~\ref{fig:VERITAS_lightcurve}. There is evidence for variability in the annual measurements. A fit of a constant to the annual flux values is poor ($\chi^2 = 9.7$, 2 degrees of freedom, P($\chi^2$) = 0.0079). This is driven by the elevated flux seen in 2018, F($>$315 GeV) = $(7.6 \pm 1.2) \times 10^{-13}$ cm$^{-2}$ s$^{-1}$, which corresponds to 0.7\% of the Crab Nebula flux \citep{Albert2008}. Although an elevated flux is seen from 3C\,264 in 2018, the observed value places it among the dimmest sources detected in the VHE band.  The monthly fluxes observed in 2018 also show evidence for VHE variability, as a similar fit of a constant is poor ($\chi^2 = 8.8$, 2 degrees of freedom, P($\chi^2$) = 0.012).  The poor $\chi^2$ comes from the factor of $2-3$ decrease in April 2018 from the elevated flux-state observed during the February to March 2018 time period.  The significance of the excess observed from 3C\,264 in 2017 and 2019 is low during each of those seasons.  Correspondingly, 99\% confidence level upper limits of F($>$315 GeV) $< 7.0 \times 10^{-13}$ cm$^{-2}$ s$^{-1}$ for 2017, and F($>$315 GeV) $< 8.2 \times 10^{-13}$ cm$^{-2}$ s$^{-1}$ for 2019, are also reported. 

\begin{figure}[tbh]
\centering
  \includegraphics[width=3.5in]{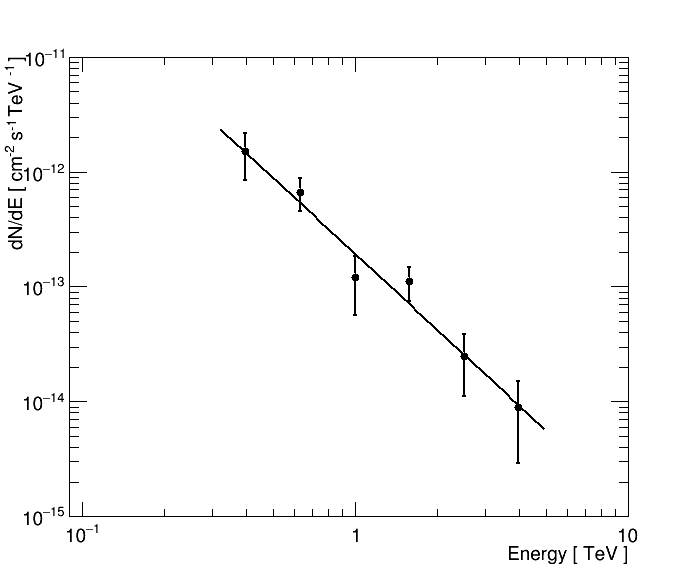} 
\caption{\label{fig:VERITAS_spectrum} VHE spectrum observed by VERITAS in 2018.}
\end{figure}

The photon spectrum from the 2018 VERITAS observations of 3C\,264 is shown in Figure~\ref{fig:VERITAS_spectrum}.  The data are well fit by a power law of the form dN/dE $\sim$ E$^{-\Gamma}$ ($\chi^2$ = 3.0, 4 degrees of freedom) with a hard photon index of $2.20 \pm 0.27_{\mathrm stat}  \pm 0.20_{\mathrm syst}$ and differential flux normalization of $(1.94 \pm 0.35_{\mathrm stat} \pm 0.58_{\mathrm syst}) \times 10^{-13}$ cm$^{-2}$ s$^{-1}$ TeV$^{-1}$ at 1 TeV.
 
The location of the VERITAS excess is determined using a two-dimensional Gaussian fit to a map of the excess of events. The centroid of the point-like excess is located (J2000) at Right Ascension (RA) $11^h 45^m 8.4^s \pm 0.7^s_{\mathrm stat}$ and declination ($\delta$) $+19^\circ 36' 29'' \pm 17''_{\mathrm stat}$. The source is accordingly named VER\,J1145+196, and it is located $0.017^\circ$ from the (ICRF) radio position of 3C\,264 of RA = $11^h 45^m 05.00903^s$ and $\delta$ = $+19^\circ 36' 22.7414''$ \citep{fey2004}. The VERITAS measurement has a systematic uncertainty of $0.007^\circ$ (25$''$), in addition to the statistical uncertainty of $0.006^\circ$. The systematic uncertainty comes largely from the accuracy of the calibration of the VERITAS pointing system, which corrects for the flexing of each telescope's optical support structure \citep{Griffiths2015}. The reconstructed source position is therefore consistent with the ICRF location at the $2\sigma$ level.

\section{Multi-Wavelength Data Sets}

%%%%%%%%%%%%%%%%%%%%%%%%%%%%%%%%%%%%%%%%%%%%%%%%%%%%%%%
% FERMI DATA
%%%%%%%%%%%%%%%%%%%%%%%%%%%%%%%%%%%%%%%%%%%%%%%%%%%%%%%
\subsection{HE $\gamma$-ray Observations}
\label{DataHE}

The Fermi Large Area Telescope (LAT) is sensitive to $\gamma$-rays from $\sim$20 MeV to $\sim$300 GeV, and it operates primarily in a survey mode that covers virtually the entire sky every few hours. 3C\,264 appears in several \emph{Fermi}-LAT catalogs and is associated with the source 4FGL\,J1144.9+1937. It is not classified as variable in the 8-year Fermi catalog \citep{4FGL2019}, where its spectrum is characterized by a power-law with a photon index of $1.94 \pm 0.10$.  Although the object is not listed as a variable source, archival \emph{Fermi}-LAT observations in the region of 3C\,264 were analyzed for the time period coincident with the VERITAS observations in 2018 (09 Feb 2018 to 21 Apr 2018), corresponding to mission-elapsed time (MET) 539827205  to 545961605, in order to probe the HE emission. A standard `unbinned' likelihood analysis was performed using the python-based \emph{Fermi} tools (ver 1.0.1). In particular, a region-of-interest (ROI) of $15^{\circ}$ around the position of 3C\,264 was analyzed including photons with energies from 100 MeV to 100 GeV. The initial maximum-likelihood optimization used a model file populated from the 3FGL catalog \citep[3FGL;][]{acero2015}. A converged fit was found after some iteration which required both adding new sources to the model file based on residual significance in the `TS' (test statistic) maps and removing some low-significance sources from the 3FGL list.  After the final XML source model file was generated, we used maximum likelihood to measure the flux and spectral index of 3C\,264 during the 2018 VERITAS observations. The spectral index has a large error, so we also ran a fit with the spectral index fixed to the 4FGL catalog value and this source flux is reported in Section~\S\ref{sec:FermiObservations}. We also generated a 95\% upper limit for the time frame of the 2017 observations using the default method of Likelihood profile fitting with the Fermi UpperLimit tool.

%%%%%%%%%%%%%%%%%%%%%%%%%%%%%%%%%%%%%%%%%%%%%%%%%%%%%%%
% X-RAY DATA
%%%%%%%%%%%%%%%%%%%%%%%%%%%%%%%%%%%%%%%%%%%%%%%%%%%%%%%
\subsection{X-ray Observations}
\label{DataXray}

\paragraph{Swift.}
\label{xrt}
The X-Ray Telescope (XRT) aboard the \textit{Neil Gehrels Swift} Observatory observed 3C\,264 (Target ID: 10512) in 2018 and 2019 for a total exposure of 54 ks. The data are almost evenly split between the two years, and these are all known XRT exposures of 3C\,264 prior to the end of the VERITAS observations in 2019. These observations were made under several target of opportunity requests, and the observation IDs are listed in Table~\ref{tab:xrtfitting} next to the appropriate spectral fits.  Since 3C\,264 has a relatively low XRT count rate, all exposures were taken in photon counting (PC) mode. The average count rate is 0.09 counts/s, well below the suggested threshold of 0.5 count/s for performing any pile-up correction.  All XRT exposures were processed and analyzed with \texttt{HEASoft V6.26.1}\footnote{https://heasarc.gsfc.nasa.gov/docs/software/heasoft/}. All level-2 data products were created locally using \texttt{xrtpipeline} V0.13.5. The spectra were extracted using \texttt{XSELECT V2.4g} and model fitting was performed using \texttt{XSPEC 12.10.1f} (via  \texttt{PyXspec}). The source region is a circle of radius, r = 20 pixels (about $45''$) while the background region has the same shape and size, and was placed nearby to avoid the single other point source in the region. 

Each XRT observation was analyzed individually, as well as grouped into the several epochs shown in Table~\ref{tab:xrtfitting}. A similar analysis was performed on each sample. Events were extracted between 0.3 and 10 keV. However, due to the relatively weak source the energy bins above 7 keV are consistent with zero flux in many cases, so model fitting was performed to obtain a flux extrapolated to 10 keV. Each spectrum was fit to a simple absorbed power law using the xspec model \texttt{phabs*powerlaw}. The hydrogen column density was fixed at $n_H = 1.96\times10^{20}\rm\ cm^{-2}$ and the remaining fit (power-law) parameters were left free. This value of $n_H$ was determined using the \texttt{nhtot} webtool\footnote{https://www.swift.ac.uk/analysis/nhtot/index.php} which facilitates the use of column density measurements described in \citet{2013MNRAS.431..394W}. Since the source is a point source, the weighted $n_H$ value was used. Past analyses with \emph{Chandra} and XMM (e.g.\ \citealp{perlman2010_3c264, Evans2006}) found that power-law models provided the best fits for 3C\,264 over those including some sort of thermal component (e.g.\ including xspec models \texttt{apec} or \texttt{bbody} to the absorbed model). This is also supported by the X-ray emission being possibly dominated by the large-scale jet, and not the core or galactic dust (see Section~\S\ref{Chandra-res}). Therefore no thermal model was included.

\paragraph{Chandra.}

3C\,264 was observed on  04 Apr 2018 with the High Resolution Camera (HRC) on board \emph{Chandra} under Director's Discretionary Time proposal 21058. The exposure time is 14.58 ks and all data analysis is conducted with CIAO version 4.11. The data is reprocessed in the standard way using the \texttt{chandra\_repro} script. The total flux from 3C\,264 is estimated using the \texttt{srcflux} script with the `wide' band appropriate for HRC. Previous \emph{Chandra} observations of 3C\,264 were made using the Advanced CCD Imaging Spectrometer (ACIS), where elongation of the source along the radio/optical jet direction was noted \citep{perlman2010_3c264}. To evaluate the extended emission in the 2018 HRC observations, the \emph{Chandra}/HRC PSF at the location of 3C\,264 was calculated using the
online \texttt{CHaRT} tool and simulations from CIAO task \texttt{psf\_project\_ray}. To reduce the statistical uncertainty (noise) in the PSF, which is matched to the data by total counts/exposure, 50 realizations of the instrument PSF from \texttt{CHaRT} 
were requested and the resulting detector-plane PSFs produced by \texttt{psf\_project\_ray} were averaged. 
This PSF is used to deconvolve the image using the Lucy-Richardson algorithm as implemented in the CIAO task \texttt{arestore}.

%%%%%%%%%%%%%%%%%%%%%%%%%%%%%%%%%%%%%%%%%%%%%%%%%%%%%%%
% OPTICAL DATA
%%%%%%%%%%%%%%%%%%%%%%%%%%%%%%%%%%%%%%%%%%%%%%%%%%%%%%%
\subsection{Optical Observations}
\label{DataOptical}

\paragraph{Ground-based}
3C\,264 was observed by two ground-based optical observatories as part of a target of opportunity campaign in 2018. Individual Johnson R band exposures were taken on eight nights between 22 Mar 2018 and 10 Apr 2018 (MJD 58199$ - $58218) at the 1.3 meter Robotically Controlled Telescope (RCT) at Kitt Peak National Observatory. In addition, Johnson V band exposures were acquired on 14 nights between 21 Mar 2018 and 20 May 2018 (MJD 58198-58258) with nearly half of the data each from two nodes of iTelescope.net: the T21 - 413mm Reflector of the New Mexico Observatory; the T27 - 770 mm Reflector of the Siding Spring Observatory; and an additional data point was taken with the T32 - 413 mm Reflector at the Siding Spring Observatory. 
The data were bias subtracted and flat-field corrected using standard IRAF routines for the RCT observations and MIRA PRO UE for the iTelescope.net observations.  V and R magnitudes were determined using differential aperture photometry with a comparison star in the same field of view as 3C 264, and a photometric radius of 10.15".

\paragraph{UVOT}

The Ultraviolet/Optical Telescope (UVOT) aboard \textit{Swift} observed 3C\,264 simultaneously during the XRT exposures described in Section~\S\ref{DataXray} (see Table~\ref{tab:xrtfitting}). Observations were made using all six filters available (v, b, u, uvw1, uvm2, uvw2), and the UVOT exposures were processed using \texttt{uvotproduct} version 2.4 to calculate the flux and generate light curves.  A circle of radius, $r = 5''$ was used for the source. The background was extracted from a nearby, source-free, circle of radius, $r = 20''$. 

\paragraph{HST}

The kpc-scale jet of 3C\,264 is visible in the optical, as well as the radio, and has been extensively observed by HST. The recent discovery of optical superluminal proper motions and colliding knots in the jet \citep{meyer2015_nature} was enabled by comparing a moderately deep ACS/WFC F606W image\footnote{HST filter information is provided here: \url{http://svo2.cab.inta-csic.es/svo/theory/fps3/index.php?mode=browse&gname=HST}} in May 2014 against earlier short WFPC2 observations from 1994, 1996, and 2002. Based on this result, a long-term monitoring campaign with HST began in 2015.  This campaign has an approximately two-year cadence following the 2014 observation; new observations were made in 2015/2016 and 2018/2019. These include polarization imaging with ACS/WFC in F606W and multi-band imaging with WFC3/UVIS for diagnostics on possible changes in the knot spectrum as the collision of knots B and C continues. Further details are in Meyer et al., (in prep.).

%%%%%%%%%%%%%%%%%%%%%%%%%%%%%%%%%%%%%%%%%%%%%%%%%%%%%%%
% RADIO DATA
%%%%%%%%%%%%%%%%%%%%%%%%%%%%%%%%%%%%%%%%%%%%%%%%%%%%%%%
\subsection{Radio Observations}
\label{DataRadio}

\paragraph{VLA}

The jet of 3C\,264 was observed by the VLA in K-band, A-configuration on 13 Aug 2015 and 02 Apr 2018.  The 2015 observation (Project 15A-507) was taken in order to provide an updated image of the jet after the discovery of the fast proper motion of two of the four knots in the optical. The April 2018 observation was obtained from Director's Discretionary Time (Project 18A-464) in response to the increased VERITAS flux observed in early 2018.  The setup and length of these observations were identical.  Therefore the 2015 epoch is an excellent reference to determine if any changes in the core or jet knots occurred during the VERITAS flare.  The data can also be compared to deep K-band imaging acquired in 1983 and 2003.

Both recent data sets were calibrated using the \texttt{CASA pipeline (version 4.7.2)}, and the scans on 3C\,264 were split off for imaging using \texttt{clean}.  Due to the wide-band observing mode (18$-$25 GHz), nterms=2 was used in \texttt{clean}.  Full polarization products were obtained after several initial rounds of self-calibration. Briggs weighting with a robust parameter of 0.5 was used for all imaging. The pixel scale was set to $0.025''$ to match the HST imaging scale. The final synthesized beam has a size of 0.12$''$x0.08$''$ and 0.18$''$x0.08$''$ in the 2015 and 2018 images, respectively. The fractional polarization and electric-vector position angle (EVPA) were calculated according to the standard formulae from the Stokes images.  

\paragraph{VLBI}

Observations with the Very Long Baseline Array (VLBA) were made on 30 Mar 2018 under a Director's Discretionary Time request related to the VHE flaring activity (Project Code BM\,450). Simultaneous multi-frequency VLBA observations were performed for a total of 4 hours at 5.0 GHz, 8.4 GHz, 12.1 GHz, and 15.3 GHz. Both circular and cross-hand polarizations were recorded with 2-bit sampling at 2048 Mbps, with 8 intermediate frequencies, each of 32 MHz bandwidth. The Los Alamos antenna did not participate due to a telecommunications problem, but useful data were obtained with the other 9 VLBA antennas.
The frequency scans were interleaved and interspersed with scans on the bright fringe calibrator source OM\,280. The total integration times were adjusted to yield a rms image noise of $\sim 0.1$ mJy/beam at each frequency.

Each frequency band was processed following standard procedures in \texttt{AIPS} and \texttt{DIFMAP}, and produced naturally weighted images with a pixel size
of 0.05 mas. The antenna polarization leakage terms were corrected using the \texttt{AIPS} task \texttt{LPCAL}.  It is not possible to calibrate the instrumental
polarization EVPA offset due to a lack of a simultaneous single-dish observation of either 3C\,264 or OM\,280. Neither source has any jet features with stable EVPA that could be used for calibration purposes. 

3C\,264 is also monitored as a part of the MOJAVE\footnote{\url{http://www.physics.purdue.edu/astro/MOJAVE/sourcepages/1142+198.shtml}} program.
In addition to the new data obtained in March 2018, MOJAVE monitoring data exists since 2016. 
The MOJAVE sample also includes an archival 15 GHz observation from 2005.
For the analysis methods of the MOJAVE program data, please refer to \cite{MOJAVE_XV}. 

%%%%%%%%%%%%%%%%%%%%%%%%%%%%%%%%%%%%%%%%%%%%%%%%%%%%%%%
% RESULTS
%%%%%%%%%%%%%%%%%%%%%%%%%%%%%%%%%%%%%%%%%%%%%%%%%%%%%%%

\section{Multi-Wavelength Results} \label{sec:Results}

\subsection{Fermi Observations} \label{sec:FermiObservations}

3C\,264 (4FGL\,J1144.9+1937) is not a particularly strong \emph{Fermi}-LAT source with an 11.4$\sigma$  significance detection in the 8-year catalog.  Correspondingly, it should only be weakly detected  ($\sim$2$\sigma$) in a few-month integration, and it is not classified as variable in the 4FGL catalog. The \emph{Fermi}-LAT data taken contemporaneous (MJD 58158$-$58229) to the VERITAS sample in 2018 indicates a higher flux F(1$-$100 GeV) = $(7.1\pm3.7) \times 10^{-9}$ ph s$^{-1}$ cm$^{-2}$ than the 4FGL catalog value of $(2.85\pm0.40) \times 10^{-10}$ ph s$^{-1}$ cm$^{-2}$. The \emph{Fermi}-LAT data taken during the main VERITAS observing period in 2018 also indicates a 
flat MeV-GeV spectrum ($\Gamma=2.1\pm0.6$), consistent with the 4FGL value ($\Gamma=1.94\pm0.10$).
Both the concurrent sample and 8-year catalog indicate the peak of the inverse-Compton component 
of the spectral energy distribution is in the GeV band.

\begin{figure}[ht]
\centering
\includegraphics[width=0.7\columnwidth, angle=270]{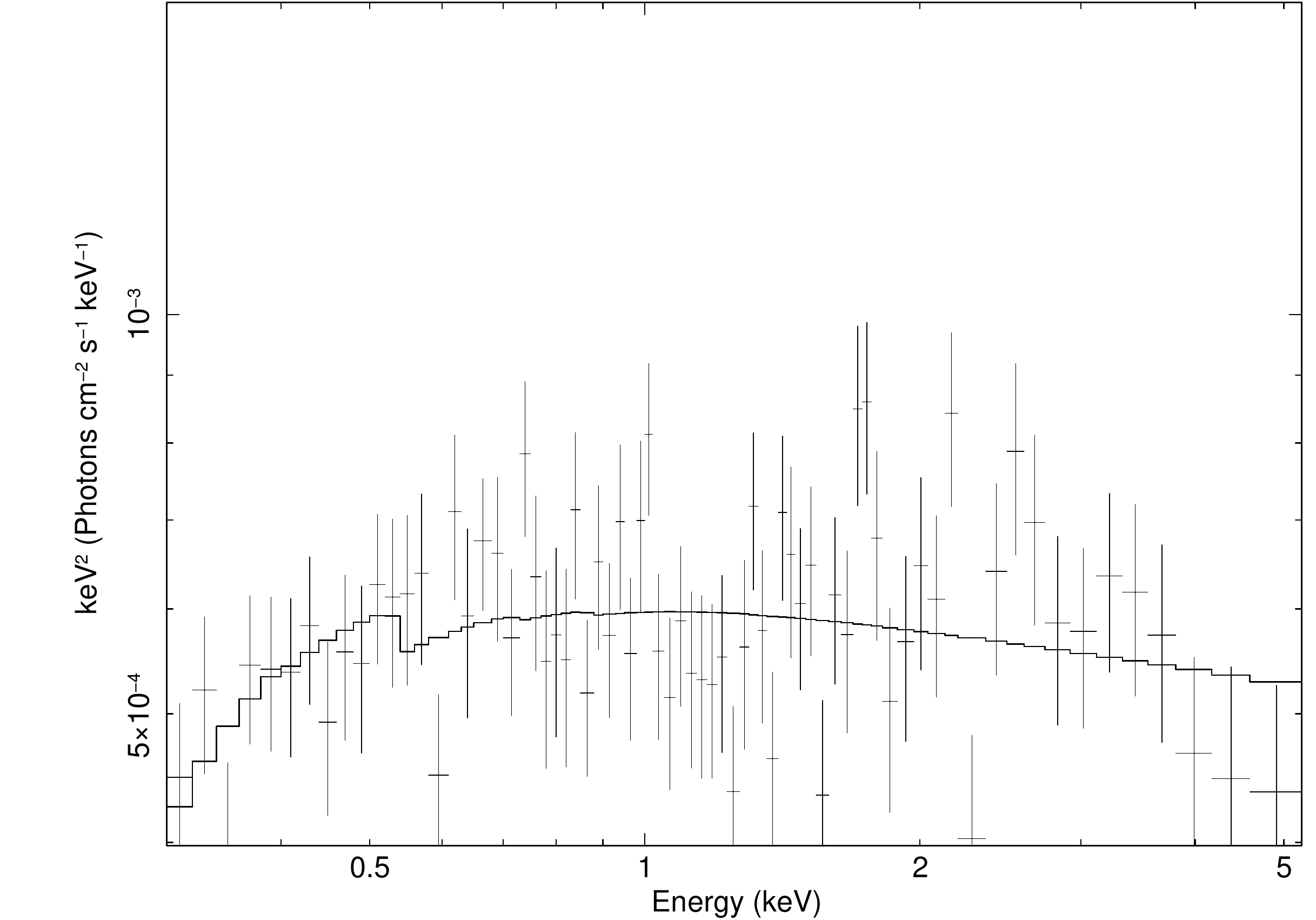}
\caption{\label{fig:swiftsed} \textit{Swift}-XRT spectrum and fit for 2018 and 2019. }
\end{figure}

\begin{table*}
\caption{\textit{Swift}-XRT spectral fit results for 3C\,264. The normalization and photon index ($\Gamma$) are the best fit results for a power-law from \texttt{phabs*powerlaw} with $n_H = 1.96\times10^{20}\rm\ cm^{-2}$. All obsid have the prefix of 000105120-.}
\centering
\begin{tabular}{cccccc}
\hline \hline
Data set & Normalization & $\Gamma$ & $\chi^2 / dof$ & 0.3-10 keV Model Flux & obsid\\
 &  [ $\rm 10^{-4}\ \rm ph\ cm^{-2}\ s^{-1}\ keV^{-1}$ ] & & &[ $10^{-12}\ \rm erg\ cm^{-2}\ s^{-1}$ ] & 000105120nn \\
\hline 
Jan 2018 & $6.54\pm0.25$ & $2.17\pm0.06$ & 317 / 486 & $3.14\pm0.12$ & 01-04 \\
Mar 2018 & $7.09\pm0.26$ & $2.06\pm0.06$ & 232 / 408 & $3.62\pm0.13$ & 06-09,11-12 \\
Apr 2018 & $5.42\pm0.27$ & $2.19\pm0.08$ & 236 / 326 & $2.58\pm0.13$ & 14,16-21 \\
%VERITAS-concurrent & $6.22\pm 0.18$ & $2.10 \pm 0.05$ & 337 / 451 & $3.10\pm0.09$ & 06-09,11,12,14,16-21\\ 
all 2018 & $7.15\pm 0.16$ & $2.10 \pm 0.04$ & 461 / 617 & $3.55\pm 0.08$ & all of the above\\
\hline
all 2019 & $4.68\pm 0.13$ & $2.20 \pm 0.05$ & 327 / 598 &  $2.22\pm 0.06$ & 22-27,29-33 \\
\hline
2018+2019 & $6.34\pm 0.12$ & $2.12 \pm 0.03$ & 487 / 549 & $3.13\pm  0.06$ & all of the above\\
\hline
\end{tabular}
\label{tab:xrtfitting}
\end{table*}

\subsection{Swift Observations}

The spectral-fit results for selected monthly and yearly epochs are shown in Table \ref{tab:xrtfitting}. The $\chi^2$ for each fit is reasonable (i.e.  $\chi^2/dof < 1$), and the 2018+2019 X-ray spectrum along with the corresponding fit is shown in Figure~\ref{fig:swiftsed} as an example. The monthly-binned X-ray flux in 2018 is significantly variable when comparing to the average with $\chi^2/dof = 32.1/2$ (P$(\chi^2) \approx 10^{-7}$). \textit{Swift} did not observe 3C~264 in Feb 2018 when VERITAS observed its highest flux, but the general trend is still apparent with the available observations from January, March, and April 2018. The XRT light-curve is shown in Figure \ref{fig:VERITAS_lightcurve}. 

Light curves of the \textit{Swift}-UVOT exposures were inspected and no time variability was found. This reinforces the expectation that the emission in this band should be dominated by stars in the galaxy and stable on this time scale. For each UVOT filter, the results were time-averaged to find a mean magnitude and energy flux. The results are shown in Table~\ref{tab:uvotflux}. 

\begin{table}[t]
\caption{\textit{Swift}-UVOT spectral information, time averaged from all 2018$-$19 \textit{Swift} observations of 3C\,264. These measurements cover a significant fraction of the entire galaxy, and not only the core or jet structure.}
\centering
\begin{tabular}{ccc}
\hline \hline
Filter & Energy Flux & Magnitude \\
 & [ $\rm  erg\ cm^{-2}\ s^{-1}$ ] & \\
\hline
v & $(3.77 \pm 0.03)\times 10^{-11}$ & $14.336 \pm 0.008$ \\
b & $(2.15 \pm 0.02)\times 10^{-11}$ & $15.305 \pm 0.008$ \\
u & $(6.87 \pm 0.06)\times 10^{-12}$ & $15.626 \pm 0.009$ \\
uvw1 &$(3.47 \pm 0.04)\times 10^{-12}$ & $16.192 \pm 0.012$ \\ 
uvm2 & $(2.61 \pm 0.04)\times 10^{-12}$ & $16.519 \pm 0.015$\\
uvw2 & $(2.49 \pm 0.02)\times 10^{-12}$ & $16.548 \pm 0.010$ \\
\hline
\end{tabular}
\label{tab:uvotflux}
\end{table}

\subsection{Ground-based Optical Observations}

No significant variability is found in the light curves from the RCT and iTelescope.net observatories.  The largest difference between any two RCT points is 0.06 magnitude, and no single iTelescope.net point differed by more than 0.16 magnitude.
The mean R-band measurement with RCT is R = 13.09 and the mean V-band measurement with iTelescope.net is V = 13.49,
which correspond to fluxes of 17.9 mJy at 640 nm and 15.4 mJy at 550 nm, respectively.  No attempt is made to subtract the host galaxy flux, and it is important to note that the integration radius for these optical data is $\sim$2 times larger than that used for the UVOT results.

\subsection{Chandra Observations}
\label{Chandra-res}

\begin{figure}[t]
\centering
\includegraphics[width=3.25in]{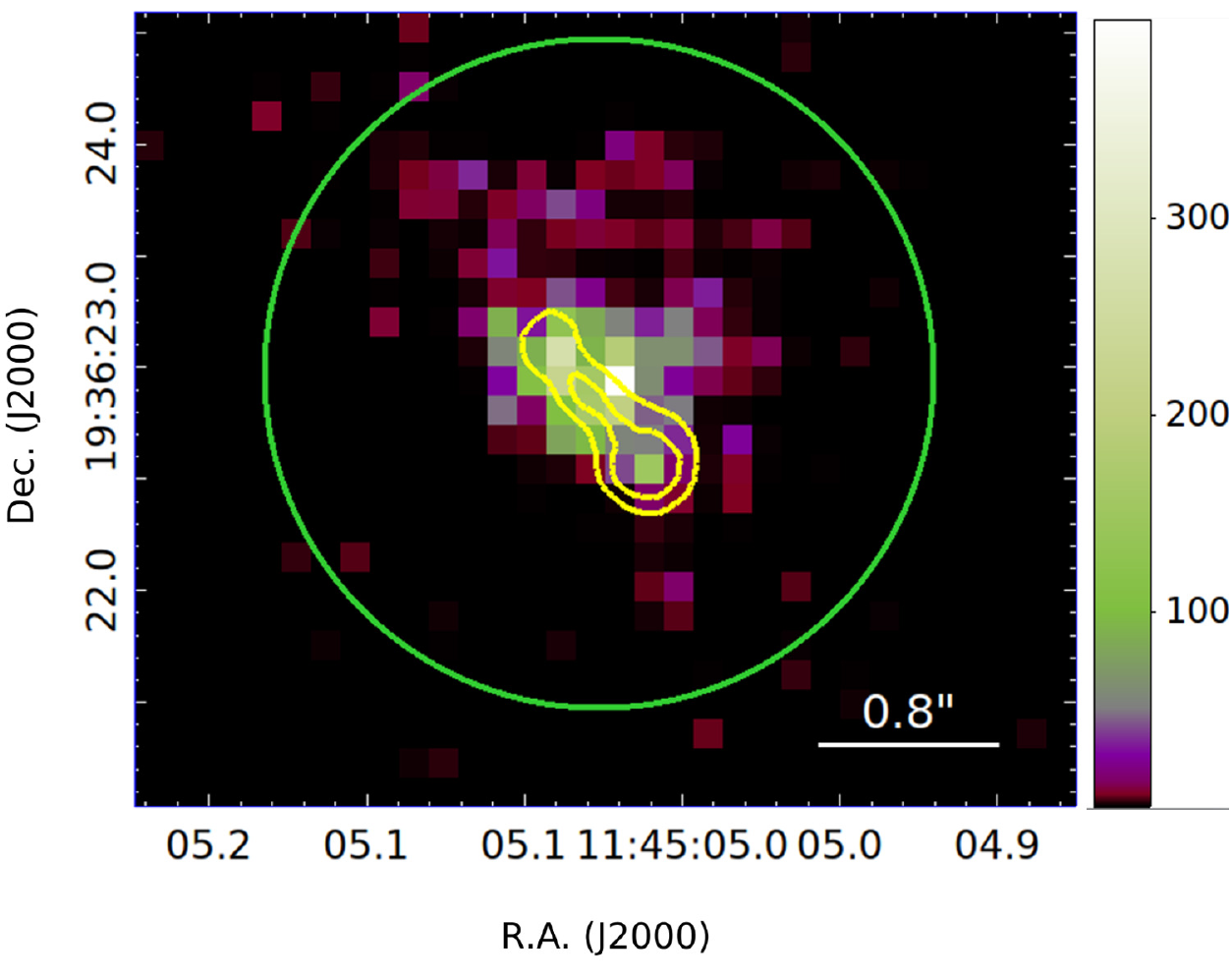}
\caption{\label{fig:chandra}  Deconvolved \emph{Chandra} HRC-I image of 3C\,264 observed on 04 Apr 2018. 
The HST image was aligned assuming the brighter component to the south is the core, and the resulting overlay of the HST contours of the jet is shown.  The previously reported thermal emission associated with the host galaxy by \cite{sun2007} is on the scales of 1.5-6$''$. This image shows very little emission on the same scales (the green circle has a radius of 1.5$''$). Color scale is in counts units.} 
\end{figure}

The deconvolved HRC-I image of 3C\,264 is shown in Figure~\ref{fig:chandra}, with the HST contours of the jet overlaid. The source is clearly extended in the image. This is also apparent from two-dimensional (2D) fitting of the (non-deconvolved) image with \texttt{sherpa}:\ a double-Sersic model fits the image best (reduced $\chi^2$ statistic of 0.042), though it leaves a residual extended flux distributed around the source. This fit model is unlikely to be physically meaningful, but the relatively large radii of the Sersic components (2.2 and 14 pixels) indicates that the bulk of the X-ray emission is extended. A point source model provides a worse fit (reduced $\chi^2$ statistic of 0.054).

The currently presented observations are the highest-resolution X-ray observations of this system to date. Previous imaging with ACIS-S and XMM suggest an extended thermal component around the AGN arising from the host galaxy on scales of $1.5-6''$ \citep[0.7-2.6 kpc,][]{sun2007}, which is considerably larger than the scale of the extended emission shown in Figure~\ref{fig:chandra}.  Indeed, outside of 1.5$''$ the 2018 observation shows very little emission. As the HRC effectively provides no spectral information, it is difficult to directly assess whether the observed extended emission could be thermal. However, \cite{perlman2010_3c264} took the `core' emission to be everything within 1.23$''$ of the peak which essentially covers the entire region of interest in Figure~\ref{fig:chandra}. Their fit to the extracted spectrum showed that a thermal component could contribute no more than 5\% of the total flux, with the rest attributed to a non-thermal power-law spectrum with a spectral index $\alpha_x$=1.24.  

Identifying the `core' location in the deconvolved \emph{Chandra} image of 3C\,264 is not unambiguous.
This is due to the absolute pointing accuracy of \emph{Chandra} (90\% uncertainty radius is 0.8$''$) and HST (typical error is $\sim$0.9$''$) and the lack of any other source in the HRC field of view.  
If the brightest pixel is assumed to be the location of the AGN core, then the brightest part of the extended and presumably non-thermal emission would be located to the south and west of the core, which seems unlikely, given the jet extends to the northwest.  Instead, if the bright component shown centered on the HST core in Figure~\ref{fig:chandra} is chosen, then the bulk of the extended emission coincides with the extended optical/radio jet rather than the core.  This is more plausible, though the brightest pixel is offset somewhat to the north side of the jet.  If this is indeed the correct identification, then the jet appears to be brighter than the core, which is unusual; the only other case where this has been observed was during the brightest outburst of HST-1 in M\,87.

Taking the total unabsorbed $0.1-10$ keV flux from the 2018 observation of $(6.91 \pm 0.2) \times 10^{-12}$ erg~cm$^{-2}~$s$^{-1}$, and adopting the spectral index $\alpha_x$=1.24 from the previous \cite{perlman2010_3c264} analysis of \emph{Chandra}/ACIS-S observations taken in 2004, gives a 1~keV monochromatic flux of 0.59$\pm$0.02\,$\mu$Jy for the entire region.  This
is approximately twice as large as the flux (0.28$\pm$0.1\,$\mu$Jy) assigned to what was referred to as the core (i.e.\ a similar region) in the 2004 observation. Crudely separating the extended jet region from the area tentatively identified as the core, we can assign approximately 80\% of the flux (470 nJy) to the extended jet. We note that the previously reported flux of 4.6$\pm$1.1 nJy for the extended jet in \cite{perlman2010_3c264} was taken for a region outside 0.8$''$ from the core, and thus from a region corresponding to the much fainter/diffuse part of the optical jet which is not detected in our observations here. The two fluxes are from different regions and should not be compared. It must be emphasized that it is not certain that the core/jet regions are correctly identified in the observations presented here due to the lack of an absolute astrometric reference. Therefore only the total X-ray flux is reported in the SEDs presented in Section~\S\ref{sec:Discussion}.

\begin{figure*}[ht]
\centering
  \includegraphics[width=0.95\linewidth]{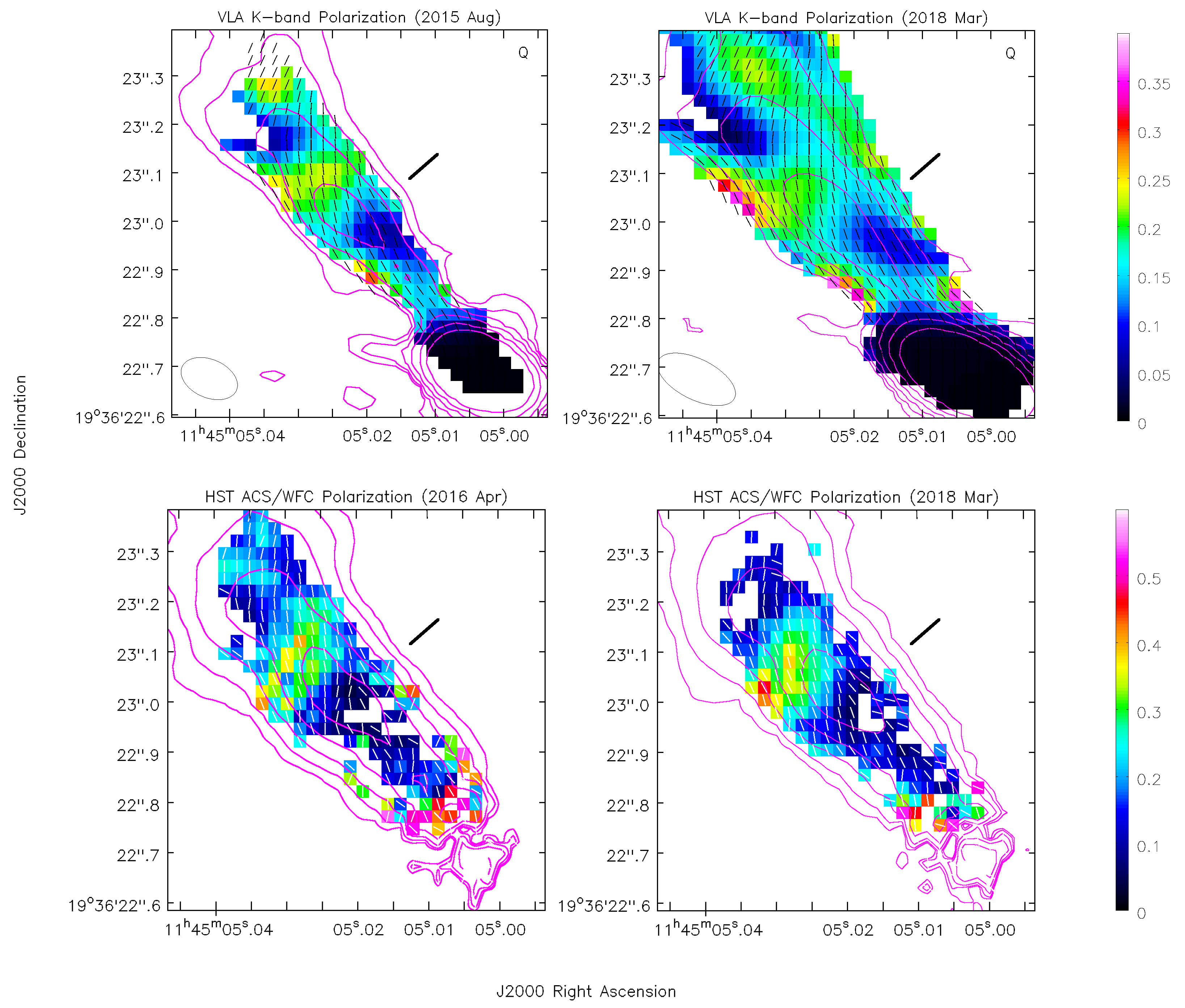}

\caption{High-resolution radio and optical polarization images of 3C\,264 from the VLA and HST. These images all show the fractional linear polarization, on a scale from 0\% to 40\% in the radio, and from 0\% to 60\% in the optical.  The VLA K-band polarization is shown in August 2015 (top left) and March 2018 (top right).  There is no striking difference in the level of polarization in the jet, both showing a peak of about 22-23\% polarization just downstream of the knot B/C collision region, indicated by the black line \citep{meyer2015_nature}.  The peak of HST-optical polarization is in the same location and reaches $\sim$15-17\%. It also shows similar levels in April 2016 (bottom left) and March 2018 (bottom right). The polarization values are uncorrected for the effect of dilution from the light from the galaxy/dust disk. In all images, the vector lines show the direction of the magnetic field (90$^\circ$ rotation from the EVPA), and the contour lines show the flux in the corresponding Stokes-I image for each epoch/band.  In the radio images, contours are drawn at 1, 2, 4, 8 and 16 times the base level of 5$\times$10$^{-4}$ Jy. In the optical images, the contour lines are drawn at 2, 4, 8, 16, and 32 times the base level of 1$\times$10$^{-8}$ Jy. }
\label{fig:polarization}
\end{figure*}

\subsection{HST Observations}

Figure~\ref{fig:polarization} shows images of the radio (VLA) and optical (HST) fractional linear polarization for the large-scale jet in 3C\, 264 (the VLA radio polarization images are described below in Section~\S\ref{VLA_pol}). In the optical, the fractional polarization images were obtained by taking the ratio of the total linear polarization to the galaxy-subtracted Stokes I image; because this subtraction removes the core, the fractional polarization shown in the core region is not meaningful. Indeed, because of the inner dust disk in 3C 264, it is difficult to disentangle the contribution of the galaxy and the synchrotron jet in the core region in the total flux images. However, assuming all the central flux in the Stokes I image comes from the synchrotron core, it suggests an optical fractional polarization at the core of 16\% and 13\% in 2016 and 2018, respectively. The integrated optical core luminosity in Stokes I (under a Gaussian fit) rose slightly between 2016 and 2018, from 185$\pm$4 $\mu$Jy to 238$\pm$5 $\mu$Jy. 

The large-scale jet shows a much higher level of polarization just downstream of the knot B/C collision zone \citep{meyer2015_nature}, as shown in Figure~\ref{fig:polarization} (the collision region is indicated with a black line). The linear polarization fraction of this feature does not appear to change significantly between April 2016 and March 2018, with a value of $\sim$25-35\% in both epochs when accounting for the contribution to the Stokes I flux from the galaxy. The size and location of this region in the two epochs is found to be consistent at approximately 0.5$''$ (220 pc) from the core and 90 pc in extent (based on a Gaussian fit).  The position angle of the magnetic field (shown in Figure~\ref{fig:polarization}, 90$^\circ$ rotated from the EVPA) also appears largely consistent between the two epochs.  It shows a smooth `flow' pattern aligned with the jet direction, with only a hint of some periodic transverse component.  Interestingly, there does appear to be a small region in the center of the knot B/C collision zone where the B-field direction becomes perpendicular to the flow.  This is consistent with the scenario outlined in \cite{meyer2015_nature} which suggests the collision is in the incipient stages. There is a possible enhancement of the linear polarization fraction which appears in the 2018 image just downstream of stationary knot A, where the polarization fraction reaches 15\% (uncorrected). However this region is very close to the bright core of the jet and differences in the orientation of HST during the two observations could change the shape and distribution of features in the Stokes I image near the core, making any features less certain.

\begin{figure*}[ht]
\centering
\includegraphics[width=6.4in]{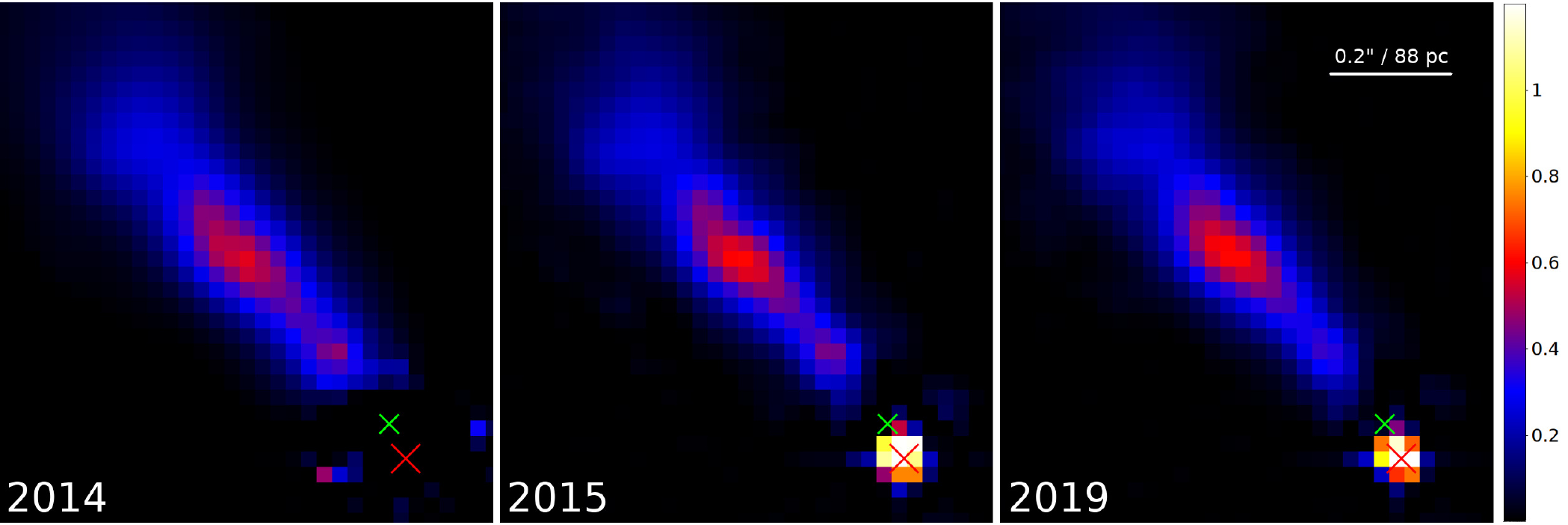}
\caption{\label{fig:HSTjet} HST images of the kpc-scale jet in 3C\,264.  In all cases, the light from the galaxy and inner dust disk is modeled and subtracted.  The images from May 2014 (left; ACS/WFC F606W), November 2015 (center; WFC3/UVIS F814W) and January 2019 (right; WFC3/UVIS F814W) are shown.  The red cross marks the location of the central black hole, and the green cross marks the location of a bend in the jet seen in VLBA imaging. Colorbar scale shown at left is in units of $\mu$Jy.} 
\end{figure*}

Figure~\ref{fig:HSTjet} shows the ACS/WFC F606W image of the jet taken in 2014 as well as the WFC3/UVIS F814W images acquired in 2015 and 2019. These observations are also useful for comparing the state of the jet before and after the increased VHE flux. The multi-band WFC3/UVIS observations taken in January 2019 were taken as replacements for June 2018 observations which missed the jet due to a problem with a gyroscope on the spacecraft. As shown, very little change can be seen between 2014 and 2019. There is a slight shift of the knot B/C centroid which is expected based on the previously detected proper motions. The change in core brightness, at 20-30\%, is typical for blazars and moderately well-aligned sources. A further discussion of the kinematics of the jet will be published in a future publication.

\subsection{VLA} \label{VLA_pol}
The VLA observations of 3C\,264 have somewhat lower resolution than the HST imaging.  However, the polarization structure also shown in Figure~\ref{fig:polarization} appears very similar.  Further, there is no obvious change between the observations taken in 2015 and those taken in 2018, during the period of increased VHE flux. The K-band core flux in 2015 was 167 mJy, and decreased to 121 mJy in 2018.

\subsection{VLBI}

\begin{figure*}[]
\centering
\includegraphics[width=6.5in]{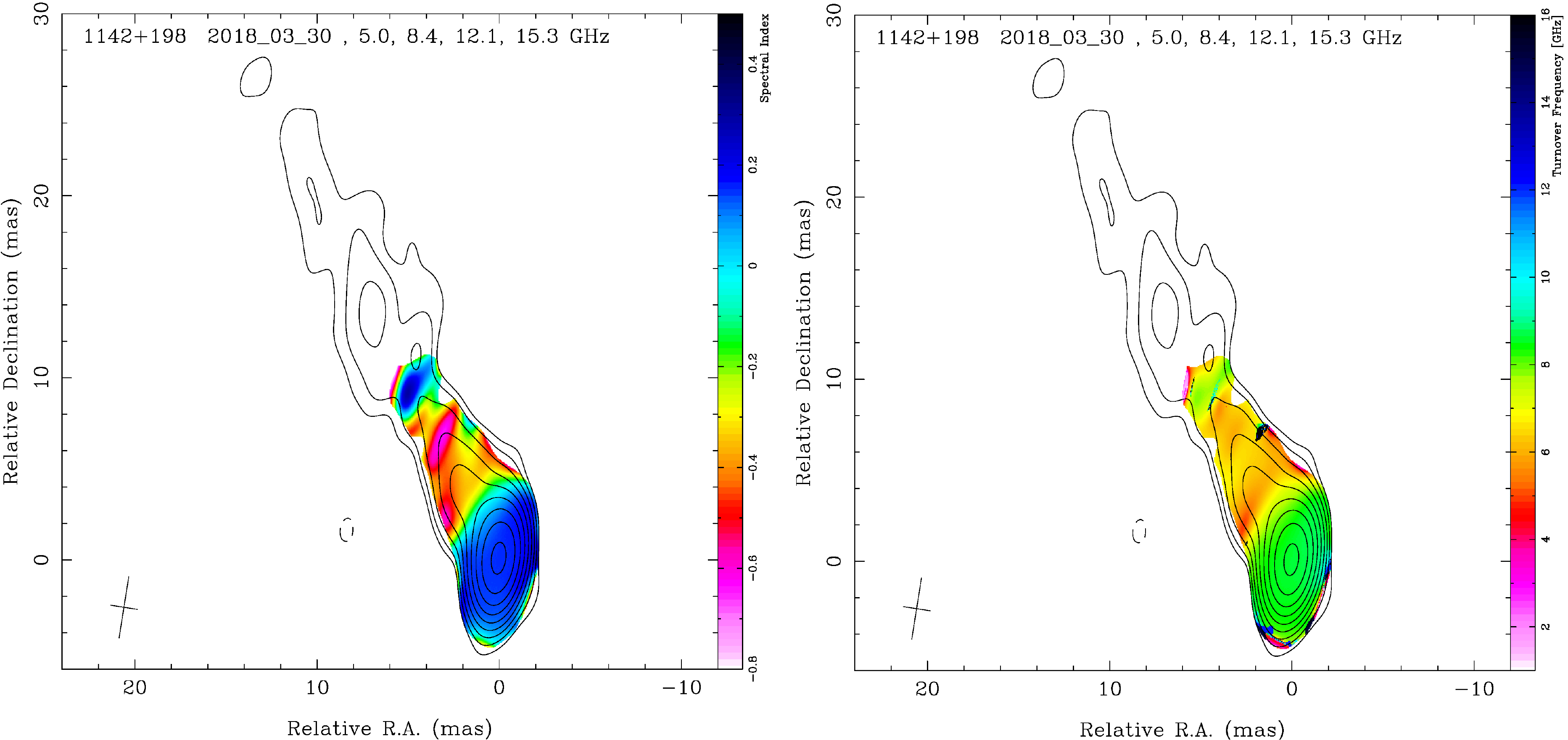}
\caption{\label{combo_spindx} Results from VLBA images of 3C\,264 taken at 5.0 GHz, 8.4 GHz, 12.1 GHz, and 15.3 GHz on 30 Mar 2018.  Maps of the radio spectral index (left) and the synchrotron spectral turnover frequency (right)
are shown. The VLBA images were restored with a common Gaussian beam having FWHM dimensions 3.4 $\times$ 1.5 mas at position angle $-9^\circ$. The 5.0 GHz total-intensity contours are drawn at successive factors of two times the base-contour level of 0.4 m\Jybm.}
\end{figure*}

After registering the VLBA images, a map of spectral index
values $\alpha$, where $S_\nu \propto \nu^{+\alpha}$, was produced by performing a
linear regression on the intensity values $S_\nu$ of each pixel.
Only pixels which exceeded 3 times the image noise level at
all four frequencies were considered. The spectral morphology map is shown
in Figure~\ref{combo_spindx}, with contours overlaid from the 5-GHz total-intensity map.

The spectral index values in Figure~\ref{combo_spindx} are only representative of
the actual spectrum in regions of the jet where the turnover frequency
does not lie between 5.0 and 15.3 GHz. To investigate this further, the
self-absorbed synchrotron spectra (see Eq.\ 4 of
\citealt{MOJAVE_XI}) were fit for each pixel and the resulting turnover
frequency values $\nu_m$ are also shown in
Figure~\ref{combo_spindx}. In this map, $\nu_m$
values below $\sim 6$ GHz and above $\sim 12$ GHz are not well
constrained by the data. However, some clear trends emerge when
comparing the two VLBA maps. The core region has a
self-absorbed spectrum peaking at $\sim 8$ GHz, and the jet becomes
optically thin roughly 4 mas (13 pc projected) downstream. At 11 mas
downstream, there is an isolated jet feature with an inverted spectrum.
The high fractional linear polarization at this location ($\sim$15\%
at 5 GHz) may be indicative of a transverse shock that is accelerating
the electrons and enhancing the magnetic field strength perpendicular
to the jet.

\begin{figure*}[]
\centering
\includegraphics[width=6.5in]{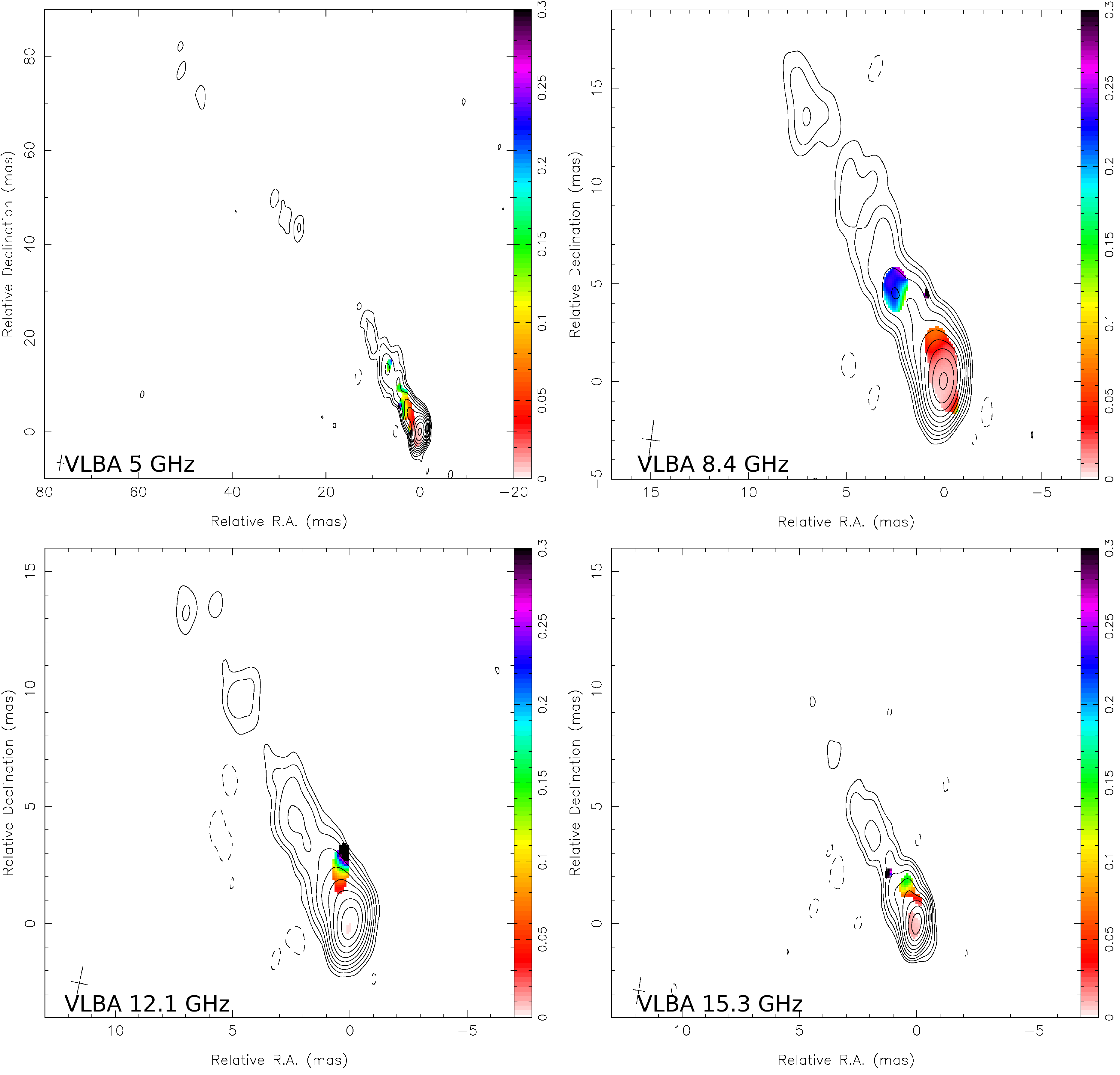}
\caption{\label{vlbi_multi} VLBA naturally-weighted contour maps of 3C 264 at 5.0, 8.4, 12.1 and 15.3 GHz. The
fractional linear polarization is overlaid in false color for pixels with
total linearly polarized intensity above 0.2, 0.45, 0.45 and 0.5 m\Jybm, respectively. The contours are drawn in successive factors of two times the base contour level of 0.2, 0.25, 0.29 and 0.3 m\Jybm. A single negative contour equal to the base contour is also drawn using dashed lines. The peak total intensity of the map is 119, 144, 132, and 116  m\Jybm, respectively. The dimensions and orientation of the restoring beam are indicated by a cross at
the lower left of each sub-figure. The 5.0-GHz image (top-left) is shown on a different, larger scale.}
\end{figure*}

The VLBA imaging of the jet is shown in Figure~\ref{vlbi_multi}. No significant change in either the core or any of the jet features were observed between any epochs taken during the VHE high state. 

%%%%%%%%%%%%%%%%%%%%%%%%%%%%%%%%%%%%%%%%%%%%%%%%%%%%%%%
% DISCUSSION
%%%%%%%%%%%%%%%%%%%%%%%%%%%%%%%%%%%%%%%%%%%%%%%%%%%%%%%
\section{Discussion} \label{sec:Discussion}
\subsection{Multiwavelength Observations}
The VERITAS observations of 3C\,264 in 2017$-$2019, as shown in Figure~\ref{fig:VERITAS_lightcurve}, indicate a period of enhanced VHE flux lasting at least several weeks in early 2018. This elevated state enabled the relatively quick discovery of the source at VHE and motivated an intensive multi-wavelength campaign to search for the origin of the VHE enhancement. However, there is no clearly identifiable source of the event. In the high-resolution radio and optical imaging from early 2018, there is no evidence of any significant change in the larger-scale jet beyond the core, i.e., no flaring event comparable to the well-known HST-1 flare in M\,87. The X-ray flux seen in the 2018 \emph{Chandra}/HRC observation is significantly increased (by a factor of 2) over that detected by \emph{Chandra}/ACIS in 2005. However, the current \emph{Chandra} imaging is inconclusive as to the location of this increase due to both the ambiguity of the core identification and the lack of a prior epoch of similar resolution. The flux observed from the core in other bands does not show a consistent pattern. It actually decreased by 27\% in the radio band between August 2015 and April 2018, while it increased by a modest 21\% in the V-band optical (F606W filter) over a similar time frame (Apr 2016 to Mar 2018). This level of optical variability appears to be typical based on observations of the core at other epochs. For example, there
is a 22\% drop in flux between the HST F475W observations in 2015 and 2019. 

\begin{figure*}[t]
\centering
\includegraphics[width=6in]{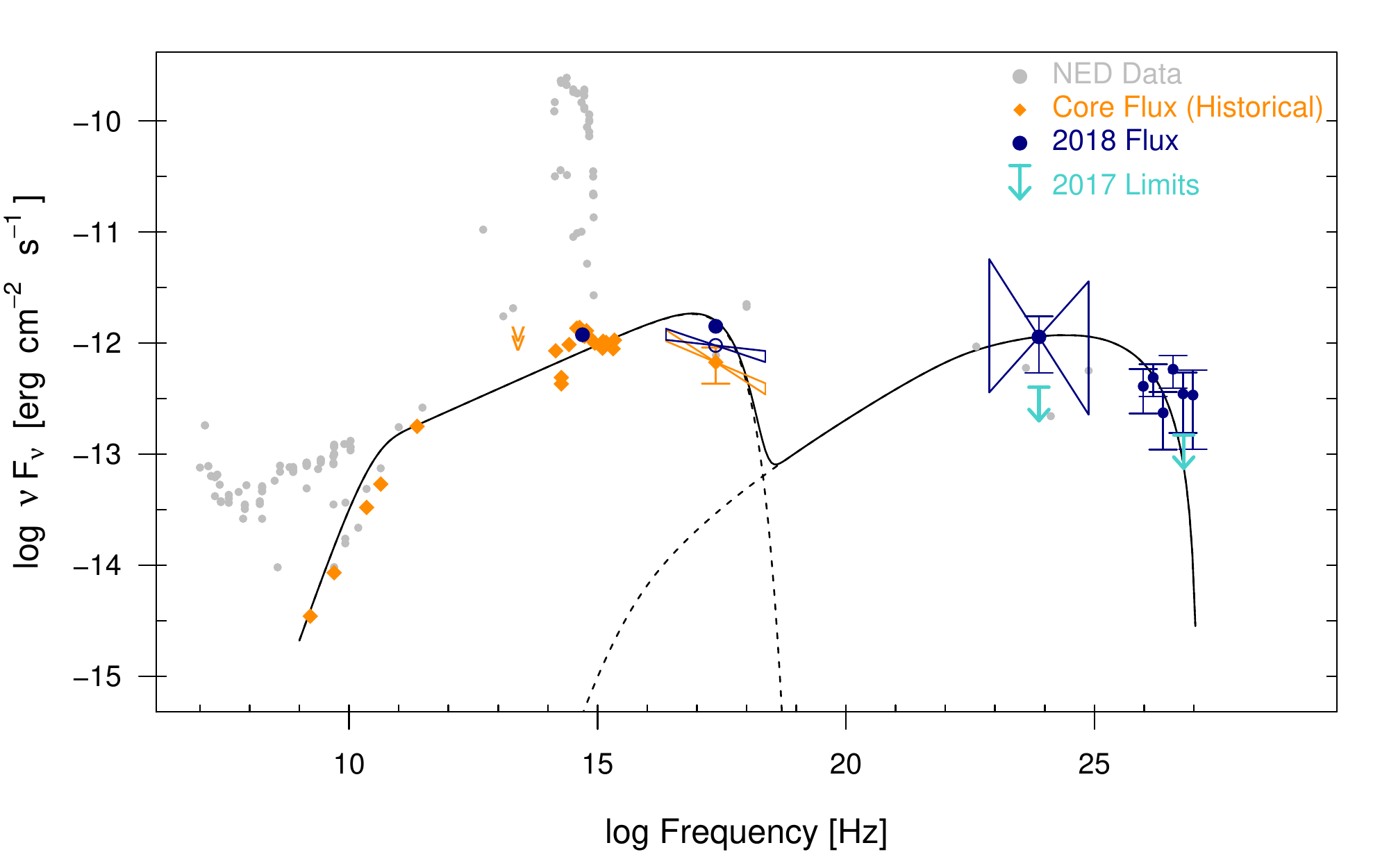}
\caption{\label{fig:mainSED} The broad-band SED for 3C\,264. Gray points are historical fluxes from NED, where the low-frequency radio is dominated by the isotropically-emitting radio lobes and the optical by the host galaxy. Shown in orange are the isolated flux values for the core as seen by VLA, ALMA, HST, and \emph{Chandra} (data taken from NED, this paper, and \citealt{perlman2010_3c264}). At high energies two temporal states are shown for 3C\,264. The cyan upper limits at GeV energies and VHE correspond to the upper limits in 2017 from \emph{Fermi}-LAT and VERITAS, while the dark blue data points and Fermi spectrum show the measurements from the 2018 enhanced state. We also show the contemporaneous optical and X-ray fluxes from 2018 as dark blue circles; in the X-rays this includes the \emph{Chandra} total measurement (filled point, no spectrum) and the VERITAS-concurrent Swift measurement (open point with butterfly spectrum). The model shown (dashed and solid lines) is a self-consistent synchrotron + SSC model with parameters typical of BL Lac objects.} 
\end{figure*}

The broad-band SED for 3C\,264 is shown in Figure~\ref{fig:mainSED} including historical core fluxes and the 2018 HST (F606W stokes I), \emph{Chandra}, Swift, \emph{Fermi} and VERITAS fluxes as well as the upper limits from $\gamma$-ray observations in 2017.
What is immediately notable about the core SED is the broadness of the lower-energy synchrotron peak, compared to typical blazars, or even M\,87. Given that only mild (factor of $2-3$) variations are seen in the 3-year VERITAS data set, and that the \emph{Fermi}-LAT flux in 2018 is only marginally higher than the 8-year average flux reported in the 4FGL catalog, it seems likely that the enhanced flux observed by VERITAS in 2018 was not related to an extreme flare (i.e., an event with 10-20$\times$ higher flux than normal) but rather a modestly elevated state.

Using a self-consistent synchrotron and SSC model (dashed and solid lines in Figure~\ref{fig:mainSED}) we are able to reasonably reproduce the observed SED. The modeling code is based on \citet{Graff}. It takes an injected electron distribution and uses a kinetic equation solved forward in time to find a steady-state electron distribution, which is then used to calculate the synchrotron and inverse-Compton emission.  Here we use a Doppler  factor of 10, and we inject a power-law electron distribution with an index of 2.6 and electron Lorentz factor confined between $200$ and $2 \times 10^6$. The comoving injected power is $ 2 \times 10^{42}$ erg/s, the comoving magnetic field $2 \times 10^{-2}$ G, and the radius of the source $2 \times 10^{16}$ cm. With these choices the source is particle dominated and the radiative cooling takes place in the slow cooling regime.  The model parameters are typical of BL Lacs, except that the Doppler factor is somewhat lower. The peak is notably at a relatively high frequency, which is unusual for radio galaxies \citep{meyer2011}. The straight portion of the synchrotron curve is not able to perfectly match the optical/X-ray flux points; this is a limitation of using a single-zone model with a power-law distribution of electron energies -- a more complex model (multi-zone and/or with a log-parabolic energy spectrum) may fit the data better, though at the cost of more input parameters. More complex modeling of this source is left to future work. Based on the single-zone model here, similar to some other BL Lacs and radio galaxies, the VHE part of the model is visually softer than the relatively flat slope indicated by the VHE data, suggesting that there is a need for multiple components or more complex models to produce harder VHE emission \citep[see e.g., the case of AP Librae,][]{Hervet,zacharias2016,petropoulou2017}.

\subsection{3C\,264 as an M\,87 Analog}

\begin{figure*}[t]
\centering
\includegraphics[width=6in]{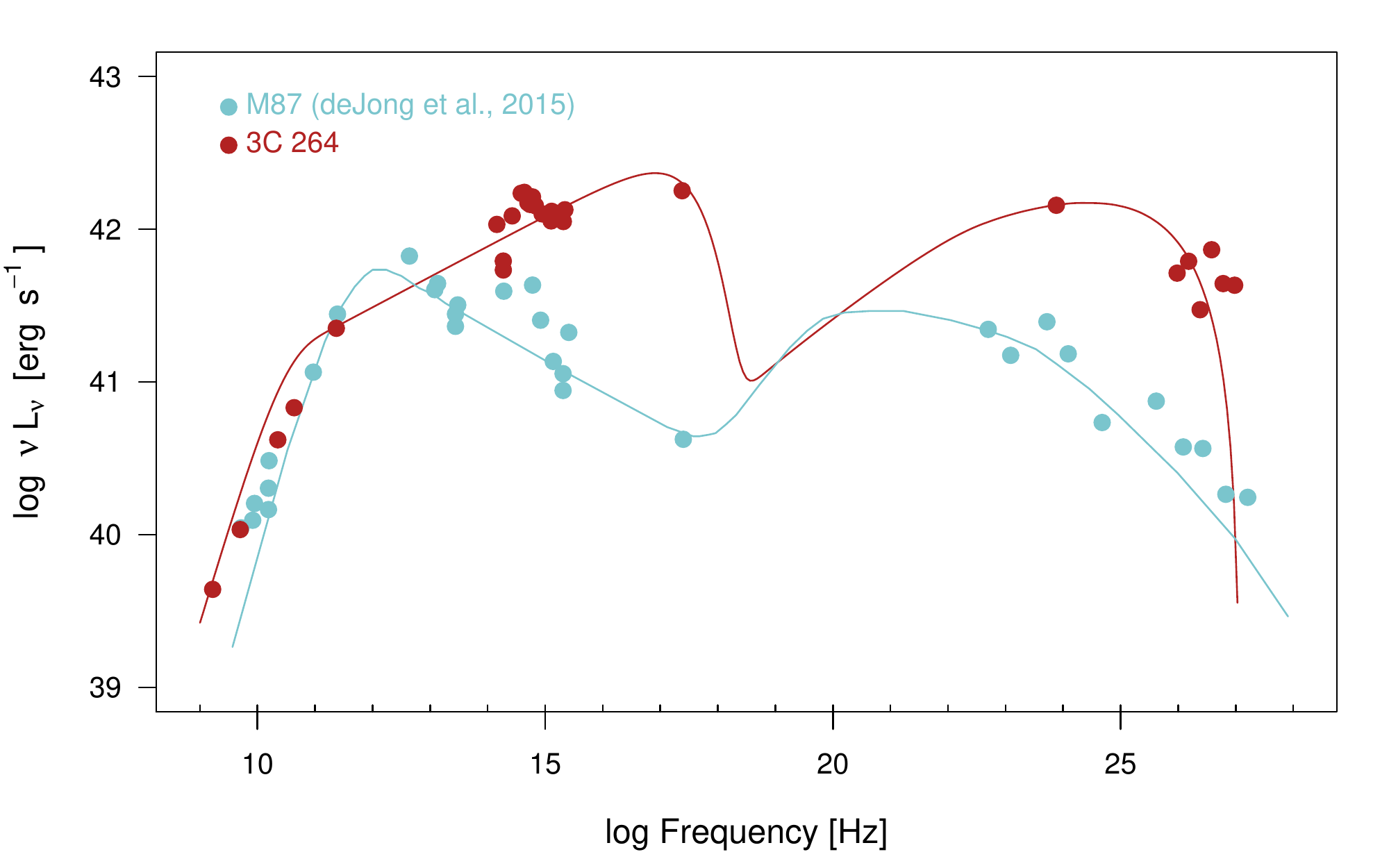}
\caption{\label{fig:compSED} SED comparison for the 3C\,264 non-thermal emission in 2018 (red) and M\,87 (light blue).  
The low-energy data points for both objects come from high-resolution imaging where the core flux can be isolated from other emission, while the HE (\emph{Fermi}-LAT) and VHE data are for the total source. For M\,87, the data and fit are taken from the `average' state SED of \citealp[][Figure 3]{dejong2015}. The model curve for 3C~264 is the same as in Figure~\ref{fig:mainSED}. While both objects have remarkably consistent radio spectra, 3C\,264 clearly has a much higher synchrotron peak, near or at the X-rays, where it is also nearly 50 times more luminous than M\,87. Similarly, the HE and VHE luminosity of 3C\,264 is also clearly higher than M\,87. }
\end{figure*}

As noted in the introduction, 3C\,264 bears some resemblance to M\,87. They have identical jet powers, exhibit a one-sided optical jet with multiple knots, and are the only objects with optical superluminal jets on kpc scales.  Each has a stationary knot feature which is the first bright optical knot in the jet, located at about 100 parsecs (projected) from the core (knot HST-1 in M\,87 and knot A in 3C\,264). Following this, both show fast superluminal motion up to $5-7c$ in the following knots with speeds decreasing along the jet \citep{meyer2013,meyer2015_nature}. The main difference is that the optical jet of 3C\,264 is about one-quarter the length of M\,87. This could be in part due to increased foreshortening due to a smaller orientation angle for 3C\,264, although 3C\,264 also has fewer optical knots than M\,87 (4 versus $\sim$7).   Note that the observed optical proper motions set the maximum angle of each jet to similar values (16$^\circ$ and 19$^\circ$ for 3C\,264 and M\,87, respectively\footnote{These angles are derived from the maximum reported speeds of $7c$ and $6c$, respectively \citep{meyer2015_nature,biretta1999}.}), but this does not necessarily mean they actually have similar orientation angles or intrinsic (as opposed to observed) speeds. 

The comparison between 3C\,264 and M\,87 is particularly interesting in light of the currently presented VHE detection because of the high-energy flaring behavior observed in M\,87 in the 2000s. These M\,87 observations consist of two distinct sets. First, \emph{Chandra} observed dramatic X-ray variability from knot HST-1 in M\,87 (100 pc from the core, projected), where the flux increased by a factor of 50 over five years \citep{harris2006_m87}, peaking in mid-2005. The dramatic increase was also seen at radio and optical wavelengths. In both the optical and X-rays, HST-1 actually outshone the core during the flare \citep{perlman2003_m87}. The knot also showed considerable shorter-timescale variability on the order of 20 days \citep{harris2009_m87}.

Secondly, during the same decade, three major VHE flares were observed from M\,87 in April 2005, February 2008, and April 2010 \citep{aharonian2006,acciari2009,aliu2012}, each with day-scale VHE variations. There has been considerable speculation about the location of these VHE flaring events.  While typically it is assumed that only the core region would be compact enough to give rise to the day-scale VHE variability, the extreme X-ray outburst of the HST-1 knot led some to consider it as an alternative site \citep{stawarz2006_m87,cheung2007,harris2009_m87,harris2012_proc}. Confirming any VHE flare arising hundreds parsecs or more downstream of the base of the jet would be a major discovery and would significantly challenge models of jet formation, especially given the required small emission regions. However, high-resolution imaging conducted at the time of the 2008 and 2010 flares seems to point to the core and not 
the HST-1 knot as being the source of the VHE flaring in M\,87 \citep{abramowski2012}.  This is based on increased core activity during these VHE flares \citep[see also][]{georganopoulos05}. While those authors make a convincing case for a `blazar-like' origin for the VHE emission in M\,87 (both during quiescent and flaring states), the VHE emission has never been conclusively shown to originate in the core or HST-1. The possibility of more than one location is also not disfavored by the data \citep{abramowski2012}. 

In light of the many points of similarity already noted between 3C\,264 and M\,87, it is interesting to compare the broad-band SEDs directly. The VHE flux of M\,87 at its peak brightness during the 2010 flare reached $\sim$10\% Crab \citep{aliu2012}. At the distance of 3C\,264 this is equivalent to 0.5\% Crab, remarkably similar to the flux detected from 3C\,264 in 2018 (0.7\% Crab). A direct comparison of the SEDs for the M\,87 core to that of 3C\,264 is shown in Figure~\ref{fig:compSED}. The data for M\,87 is taken from \cite{dejong2015} and represents an average state for the source, while the data and models for 3C~264 are the same as in Figure~\ref{fig:mainSED}. Here for both sources the isolated core measurements are used from radio to X-rays while total luminosity results (presumed to be dominated by the core) are reported at HE and VHE. It is interesting that the radio portions of the SEDs of 3C\,264 and M\,87 are practically identical,
but then deviate from each other at frequencies above $\sim 10^{13}$ Hz. The 3C\,264 synchrotron spectrum peaks somewhere between the optical and the X-rays, and the M\,87 synchrotron spectrum peaks somewhere around $\sim 10^{13}-10^{14} $ Hz. The high-energy SED of 3C\,264 is about 10 times brighter than that of M\,87. This behavior can be explained in the context of models with velocity profiles, such as a decelerating flow \citep{georganopoulos03} or a fast spine-slow sheath jet \citep{ghisellini2005} where the two jets are physically similar, but have different orientations, with 3C\,264 being closer to the line of sight than M\,87. 
In such a scenario, where the high-energy electrons produce the optical to X-ray synchrotron emission and the $\gamma$-ray inverse-Compton emission comes from the faster parts of the flow, misaligning the jet causes the more highly beamed emission to correspondingly drop faster as the jet moves away from the observer. 
Qualitatively, this would produce something like the observed differences between the two SEDs in Figure~\ref{fig:compSED}. Detailed modeling work to test this scenario will be considered in a future publication.  

\section{Summary}

VHE $\gamma$-ray emission was discovered from the radio galaxy 3C\,264 by VERITAS in the spring of 2018. This AGN
is the most distant radio galaxy detected in the VHE to date, and the discovery was facilitated by a period of enhanced VHE flux lasting for several weeks.  An extensive suite of contemporaneous multi-wavelength observations was acquired
to probe the underlying emission mechanism. These include high-resolution observations with the VLBA, VLA, HST and \emph{Chandra}, as well as observations by \emph{Swift} in the optical and X-ray, $\gamma$-rays by the \emph{Fermi}-LAT and ground-based optical observations. The mild VHE variability observed by VERITAS in 2017$-$2019 suggests that 3C\,264 did not experience a strong flare, but rather a period of modestly enhanced flux. The source of this enhanced flux is most likely the unresolved core, based on the lack of any notable change in any of the high-resolution \emph{Chandra} or HST imaging compared with previous epochs spanning the last decade; we also did not observe any large changes in the core flux at lower frequencies.
A qualitative inspection of the SED for the jet of 3C\,264 shows it is somewhat unusual for a radio galaxy, with a relatively high-frequency synchrotron peak near the X-rays. 3C\,264 could be
considered a more distant analog of the well-studied VHE source M\,87 based on both its beamed and unbeamed radio emission and its kinematic profile. If it is intrinsically similar, then 3C\,264 is likely oriented at a 
smaller angle to the line-of-sight. 

\acknowledgments
This research is supported by grants from the U.S. Department of Energy Office of Science, the U.S. National Science Foundation and the Smithsonian Institution, and by NSERC in Canada. This research used resources provided by the Open Science Grid, which is supported by the National Science Foundation and the U.S. Department of Energy's Office of Science, and resources of the National Energy Research Scientific Computing Center (NERSC), a U.S. Department of Energy Office of Science User Facility operated under Contract No. DE-AC02-05CH11231. We acknowledge the excellent work of the technical support staff at the Fred Lawrence Whipple Observatory and at the collaborating institutions in the construction and operation of the instrument. E.T. Meyer acknowledges the support of HST grant GO-14159.

\clearpage
\bibliography{references} %expects file references.bib



\end{document}